\documentclass[eqsecnum,showpacs,pre,twocolumn]{revtex4}
\usepackage{graphicx}
\usepackage{amsmath}
\usepackage{bm}
\usepackage{epsfig}
\usepackage{amsfonts}
\usepackage{amssymb}
\usepackage[usenames]{color}
\usepackage[latin1]{inputenc}
\begin{document}

\newcommand{\brm}[1]{\bm{{\rm #1}}}
\newcommand{\Ochange}[1]{{\color{red}{#1}}}
\newcommand{\Ocomment}[1]{{\color{PineGreen}{#1}}}
\newcommand{\Hcomment}[1]{{\color{ProcessBlue}{#1}}}
\newcommand{\Hchange}[1]{{\color{BurntOrange}{#1}}}

\title{Distribution functions in percolation problems}
\author{Hans-Karl Janssen}
\affiliation{Institut f\"ur Theoretische Physik III, Heinrich-Heine-Universit\"at, 40225
D\"usseldorf, Germany}
\author{Olaf Stenull}
\affiliation{Department of Physics and Astronomy, University of Pennsylvania, Philadelphia
PA 19104, USA}
\date{\today}

\begin{abstract}
\noindent
Percolation clusters are random fractals whose geometrical and transport properties can be characterized with the help of probability distribution functions. Using renormalized field theory, we determine the asymptotic form of various of such distribution functions in the limits where certain scaling variables become small or large. Our study includes the pair-connection probability, the distributions of the fractal masses of the backbone, the red bonds and the shortest, the longest and the average self-avoiding walk between any two points on a cluster, as well as the distribution of the total resistance in the random resistor network. Our analysis draws solely on general, structural features of the underlying diagrammatic perturbation theory, and hence our main results are valid to arbitrary loop order.
\end{abstract}
\maketitle

\section{Introduction and summary of results}
Percolation~\cite{StAh94} clusters are random fractals. Due to their randomness, it is natural to characterize their properties, such as geometrical and transport properties, by means of probability distribution functions. The perhaps most basic of these distribution functions is the the percolation correlation function $\chi (\brm{r} - \brm{r}^\prime)$ which measures the pair connection probability, i.e., the probability that two sites located at space coordinates $\brm{r}$ and $\brm{r}^\prime$ are connected.
For many applications, it is useful to consider percolation models that allow for additional processes to unfold on the clusters. The most prominent model of this kind is the random resistor network (RRN) where  occupied bonds are viewed as resistors so that the clusters can transport electric currents. One of the typical questions that are addressed when studying the RRN is that of the total resistance $R (\brm{r},\brm{r}^\prime)$ between 2 leads located at points $\brm{r}$ and $\brm{r}^\prime$ {\em on the same cluster}. Because the underlying cluster is random, the $R (\brm{r},\brm{r}^\prime)$ is a random number governed by a distribution function, $P (R, \brm{r} - \brm{r}^\prime)$. Experimentally, this distribution might be determined as follows: Measure the resistance between all pairs of sites with separation $|\brm{r} - \brm{r}^\prime|$ using an ohmmeter. Discard all measurements yielding an infinite value as these measurements probe sites belonging to different clusters. Prepare a histogram that collects the measured finite values. Then, $P (R, \brm{r} - \brm{r}^\prime) dR$ corresponds to the number of points in the histogram between $R$ and $R+dR$ divided by the number of measurements with finite outcome. A similar experiment has been performed, e.g., some 20 years ago by Rammal, Lemieux and Tremblay~\cite{RamLemTem_85}.

Other interesting measurable quantities that are intimately related to the transport properties of percolation clusters include the masses (the number of bonds) of various fractal substructures of the clusters. The most prominent substructures of a cluster are perhaps the backbone (the set of bonds that carry current), the red bonds (the bonds that are singly connected and hence carry the full current) as well as the shortest, the longest and the average self-avoiding walk (SAW) on the cluster. Their masses are random numbers, and the corresponding distribution functions can be measured in much the same way as described above for the total resistance.

In this paper, we study the distribution functions of the pair connection probability, the total resistance and the masses of fractal substructures by using renormalized field theory. Our work is based on the Harris-Lubensky (HL) model~\cite{harris_lubensky_84,harris_kim_lubensky_84,harris_lubensky_87b} for the RRN and its nonlinear generalization by Harris~\cite{harris_87}. In this generalized model, the bonds are assumed to have a nonlinear current-voltage characteristic of the type $V \sim I^r$~\cite{kenkel_straley_82}.  The main benefit of this generalization is that it allows to map out the aforementioned fractal substructures by taking the appropriate limits with respect to $r$: the limit $r\to -1^+$ leads to the backbone, whereas $r\to \infty$, $r\to 0^+$ and $r\to 0^-$ lead to the red bonds, the shortest and the longest self-avoiding path, respectively. Of course, one retrieves the linear RRN for $r\to 1$. Via these limits, the distribution functions of the total resistance and the fractal masses are all related to one single distribution function, namely that of the total nonlinear resistance. Regarding the average SAW, the story is somewhat more complex. Obviously, the length of the average SAW lies between the lengths of the shortest and the longest SAW. In a certain sense, the average SAW is concealed somewhere in the discontinuity at $r=0$, and it not clear how to extract its length by taking a limit of the nonlinear RRN. Recently~\cite{janssen_stenull_SAW_2007}, we proposed a method to calculate the scaling form of this length via using our real-world interpretation of the Feynman diagrams of the RRN, which we will explain below. In this paper, we use this method (which is intimately related to a correct interpretation of the Meir-Harris model~\cite{MeHa89}) to determine the distribution of this length.

In their seminal paper~\cite{harris_lubensky_87b}, HL studied among many other interesting things the asymptotic form at criticality of the distribution $P (R, \brm{r})$ of the linear total resistance for large $R$. Their calculation is based on the assumption that the {\em unconditional} probability $P_{\text{uc}} (R, \brm{r})$ for measuring $R$, which differs from $P (R, \brm{r})$ in that infinite resistances are not discarded, becomes independent of $\brm{r}$ for large $R$. This assumption is problematic, however, because large $R$ corresponds due to scaling to short distances, and in critical phenomena, short distances generically lead to extra singularities which has qualitative consequences~\cite{amit_zinn-justin}. Our careful analysis, based on the so-called short-distance expansion as a field-theoretic tool, reveals that this assumption breaks down beyond mean-field theory, and hence the long-standing result by HL turns out being incorrect for dimensions $d\leq 6$.

Before moving into the depths of field theory, we would like to summarize for the convenience of the reader the resulting picture of the various distribution functions. It is well established, that the pair connection probability obeys the scaling form
\begin{equation}
\chi(\mathbf{r},\tau)=  |\mathbf{r}|^{2-d-\eta}\,  \hat{\chi}\big(\tau |\brm{r}|^{1/\nu}\big)\, ,
\end{equation}
where $\tau$ is the control parameter that measures the deviation from the percolation point. In bond percolation, e.g.,  $\tau$ is proportional to $(p_c -p)$, where $p$ is the bond occupation probability, and $p_c$ is its critical value. $\eta$ and $\nu$ are the order parameter and correlation length exponents for isotropic percolation, respectively. $\hat{\chi} (x)$ is a scaling function, whose asymptotic behavior we calculate for small and large $x$. For small $x$, we find
\begin{equation}
\hat{\chi}(x) = A_0 +A_{1}x+A_{2}x^{2}+A_{3}x^{d\nu-1}+\cdots\,,
\end{equation}
where $A_0$ and so on are expansion coefficients. For large $x$, we obtain
\begin{equation}
\hat{\chi}(x)  = A\, x^{(d-3+2\eta)\nu/2}\, \mathrm{\exp}\left(  -ax^{\nu}\right) + \cdots \, ,
\end{equation}
with constants $A$ and $a$. The distribution $P_r(R,\mathbf{r})$ of the total nonlinear resistance obeys at criticality, $\tau =0$, the scaling form
\begin{align}
\label{scaleFromPrCrit}
P_r(R,\mathbf{r})   =R^{-1}\, \Omega_r\big(R\left\vert \mathbf{r}\right\vert ^{-\phi_r/\nu}\big) ,
\end{align}
where $\phi_r$ is the resistance exponent of the nonlinear RRN. As discussed above, the distribution function for the total linear resistance is implied in this result for $r\to 1$, that of the backbone mass for $r\to -1^+$ and so on.  For large $s$, we find that the scaling function $\Omega_r (s)$ behaves as
\begin{align}
\label{OmegaTlargeS}
\Omega_r (s) &=  C_{r,1}\, s^{-(d\nu-1)/\phi_r}+C_{r,2}\, s^{-d\nu/\phi_r}
\nonumber\\
&+C_{r,3}\, s^{-(d\nu+\phi_r-1)/\phi_r}+C_{r,4}\, s^{-(d+\bar{\omega})\nu/\phi_r} + \cdots,
\end{align}
where $C_{1,r}$ and so on are constants, and where $\bar{\omega}$ is the so-called Wegner exponent. For small $s$, we get
\begin{align}
\label{OmegaTsmallS}
\Omega_r (s) &=  C_{r} \, s^{-(d-2+2\eta)\nu/2(\phi_r-\nu)}\exp\left(  -c_r s^{-\nu/(\phi_r-\nu)}\right)
\nonumber\\
& + \cdots
 ,
\end{align}
 with constants $C_r$ and $c_r$. Equations~(\ref{scaleFromPrCrit}) to (\ref{OmegaTsmallS}) also describe the distribution of the length of the average SAW; we merely have to replace $\phi_r$ by $\phi_0 := \nu/\nu_\text{SAW}$, where  $\nu_\text{SAW}$ is the usual scaling exponent of the average SAW on a percolation cluster, and we also have to replace nonuniversal constants $C_{r,1}$ and so on by other nonuniversal constants, say $C_{0,1}$ and so on. We note that similar asymptotic laws were derived a long time go by des\,Cloiseaux \cite{dCl74}, Fisher \cite{Fi66}, and McKennzie and Moore \cite{McKMo71} for SAWs on isotropic non-disordered substrates.

The remainder of this paper has the following outline: Section~\ref{reviewPottsRRN} briefly reviews the field theoretic formulation of the Potts model and the HL model and its nonlinear generalization to provide background information and to establish notation. Section~\ref{representations} treats the renormalization of composite fields that play a role in the short distance expansion. Section~\ref{shortDistanceExpansion} contains the core of our short distance expansion. Section~\ref{shapeDistFuncts} presents the derivation of our final results for the distribution functions. Section~\ref{concludingRemarks} gives a few concluding remarks.

\section{Field-theoretic Percolation Models Based on the Potts model -- a Brief Review}
\label{reviewPottsRRN}
There are two main field-theoretic approaches for percolation problems: one is based on the so-called generalized epidemic process, and the other is based on the $q$-state Potts model in the limit $q\to 0$. The latter is intimately related to the RRN and thus it is particularly useful for studying transport properties of percolation clusters. In this paper we focus on the Potts model and the RRN along with its nonlinear generalization. Here, in this section we review some aspects of the field theory of the Potts model and the RRN that are essential for our present work, in order to provide the reader with background information and to establish notation.

\subsection{Potts model}

\subsubsection{Hamiltonian}
In the literature, the minimal model or field theoretic Hamiltonian for the $q=(n+1)$-state Potts model has been written in several forms. The perhaps simplest form
\begin{equation}
\label{PottsHamiltonian}
\mathcal{H}=\int d^{d}r\,\sum_{k=0}^{n}\left\{  \frac{1}{2}\left(
\nabla\varphi_{k}\right)  ^{2}+\frac{\tau}{2}\varphi_{k}^{2}+\frac{g\sqrt{n+1}}
{6}\varphi_{k}^{3}\right\} ,
\end{equation}
is achieved if the $(n+1)$-component order parameter field $\varphi_k$ satisfies the constraint
\begin{equation}
\label{PottsConstraintOriginal}
\sum_{k=0}^{n}\varphi_{k}=0\,,
\end{equation}
because, in this case, $\varphi_k$ shows most directly the underlying symmetry of the Potts model that arises from the equivalence of the $n+1$ states. The order parameter with constraint (\ref{PottsConstraintOriginal}) constitutes an irreducible representation of the group of permutations of $n+1$ elements, i.e., the symmetric group $S_{n+1}$. In group theoretical language \cite{Ham62} this is the representation $(n,1)$ \cite{WaYo78}. The parameter $\tau$ measures the distance from the critical point in the phase diagram, which is given in mean field theory by $\tau_c = 0$. Note that there is only a single expansion coefficient or coupling constant $g$ decorating the cubic term since the third order direct product $(n,1)^3$ contains the identity representation only once. Hence, there is just one cubic invariant of the order parameter. For the model to describe (purely geometric) percolation, one has to take the limit $n\to 0$.

For addressing certain questions, it can be useful to switch from $\varphi_k$ to its Fourier transform $\tilde{\varphi}_{\lambda}$ and vice versa via the relations
\begin{subequations}
\begin{align}
\label{pottsFourierTrafo1}
\varphi_{k}  &  =\frac{1}{\sqrt{n+1}}\left.  \sum_{\lambda}\right.  ^{\prime
}\mathrm{e}^{ik\lambda}\tilde{\varphi}_{\lambda}\,,
\\
\label{pottsFourierTrafo2}
\tilde{\varphi}_{\lambda}  &  =\frac{1}{\sqrt{n+1}}\sum_{k=0}^{n}%
\mathrm{e}^{-ik\lambda}\varphi_{k}\,,
\end{align}
\end{subequations}
with $\lambda$ given by
\begin{equation}
\lambda=\frac{2\pi l}{n+1}\operatorname{mod}2\pi\qquad\text{\textrm{with}%
}\qquad l=0,1,\ldots,n \, .
\end{equation}
The prime in Eq.~(\ref{pottsFourierTrafo1}) indicates that $\lambda = 0$ is excluded from the sum over $\lambda$ due to the constraint~(\ref{PottsConstraintOriginal}). Alternatively, one can admit $\lambda = 0$ if one demands that $\tilde{\varphi}_{\lambda}$ satisfies the constraint $\tilde{\varphi}_{\lambda = 0} = 0$  that is a consequence of Eq.~(\ref{PottsConstraintOriginal}). In terms of $\tilde{\varphi}_{\lambda}$, the field theoretic Hamiltonian for the Potts model reads
\begin{align}
\mathcal{H}&=\int d^{d}r \Bigg\{  \frac{1}{2}\sum_{\lambda}\tilde{\varphi}_{-\lambda}\left[  \tau -\nabla^{2}\right]  \tilde{\varphi}_{\lambda}
\nonumber \\
&+\frac{g}{6}\sum_{\lambda,\lambda^{\prime}}
\tilde{\varphi}_{\lambda}\tilde{\varphi}_{\lambda^{\prime}}\tilde{\varphi}_{-\lambda-\lambda^{\prime}}\Bigg\}
\,.
\end{align}

\subsubsection{Renormalization}
Throughout the rest of this paper, we will draw heavily on the well established methods of renormalized field theory. We will outline the steps that we take as much as possible. Given the space-constraints of this type of paper, however, we have to refer to the textbooks on field theory~\cite{amit_zinn-justin} for general background and technical details.

At and below the upper critical dimension $d_c = 6$, mean field theory, i.e., saddlepoint-integration of functional integrals followed by naive perturbation expansions (with a physical finite cutoff),  breaks down. Naive perturbation expansions produce IR-divergencies near the critical point, and renormalization group methods must be employed instead to sum up the leading divergencies of the complete expansion. In diagrammatic perturbation calculations using dimensional regularization instead of a finite cutoff, poles in $\varepsilon = 6-d$ arise. These $\varepsilon$-poles can be eliminated from the theory by employing the renormalization scheme
\begin{subequations}
\label{reno}
\begin{align}
\varphi \rightarrow \mathring{\varphi}&=Z^{1/2}\varphi\, ,  \\
\tau \rightarrow \mathring{\tau }&=Z^{-1}Z_{\tau }\tau +\mathring{\tau}_c \, ,\\
g \rightarrow \mathring{g}&= Z^{-3/2}Z_{u}^{1/2} g, \\
u & = G_{\varepsilon } \mu^{-\varepsilon} g^2 ,
\end{align}
\end{subequations}
where the ring  $\mathring{}$  denotes bare, unrenormalized quantities. $u$ is an effective coupling constant that appears naturally in any diagrammatic calculation based on the Potts Hamiltonian~(\ref{PottsHamiltonian}). $\mu$ is an inverse length-scale, and the factor $\mu^{-\varepsilon}$ makes $u$ dimensionless in the sense that its engineering dimension vanishes. $G_{\varepsilon }=\Gamma (1+\varepsilon /2)/(4\pi )^{d/2}$ is a convenient numerical factor depending on space dimension. The renormalization factors $Z$, $Z_\tau$ and $Z_u$ are known to three-loop order~\cite{alcantara_80}. As far as explicit diagrammatic calculations are concerned, we will not go in the present paper beyond two-loop order, to which the renormalization factors are given by
\begin{subequations}
\begin{align}
Z  &  =1-\left(  n-1\right)  \frac{u}{6\varepsilon}
\nonumber \\
&+\left(  n-1\right)
\left(  \frac{5n-11}{\varepsilon}-\frac{13n-37}{12}\right)  \frac{u^{2}%
}{36\varepsilon},\\
\label{2loopResTau}
Z_{\tau}  &  =1-\left(  n-1\right)  \frac{u}{\varepsilon}+\left(  n-1\right)
\left(  \frac{5n-9}{\varepsilon}-\frac{23n-47}{12}\right)  \frac{u^{2}%
}{4\varepsilon},\\
Z_{u}  &  =1-\left(  n-2\right)  \frac{2u}{\varepsilon}
\nonumber \\
&+\left(  \frac
{7n^{2}-29n+30}{\varepsilon}-\frac{23n^{2}-99n+118}{12}\right)  \frac{u^{2}%
}{2\varepsilon},
\end{align}
\end{subequations}
if one uses minimal subtraction, as we do. The critical value of the control parameter $\mathring{\tau}_c=\mathring{g}^{4/\varepsilon}S(\varepsilon)$, where the Symanzik-function $S(\varepsilon)$ contains simple IR-poles at all values $\varepsilon=2/k$ with $k=1,2,...$, becomes formally zero in the $\varepsilon$-expansion. Note that the renormalization factors are at any order in perturbation theory polynomials in $n$, and hence it is guaranteed that they are well behaved in the limit $n\to 0$.

\subsection{Nonlinear random resistor network}
\label{reviewNRRN}

In their seminal work on linear RRN, HL showed, based on earlier ideas by Stephen~\cite{stephen_78}, that the field theoretic Hamiltonian for the Potts model can be extended to describe the RRN. We refer to their field theoretic model as the HL model. Shortly after the HL model was introduced, Harris~\cite{harris_87} took up ideas by Kenkel and Straley~\cite{kenkel_straley_82} and generalized the HL model so that it captures nonlinear voltage-current characteristics of the type $V \sim I^r$. More precisely speaking, this nonlinear RRN features bonds between nearest neighboring sites $i$ and $j$ that obey a generalized Ohm's law $V_j - V_i = \rho_{i,j} I_{i,j} | I_{i,j}|^{r-1}$, where $\rho_{i,j}$ is the nonlinear bond-resistance. As is customary, we use here and in the following the symbol $r$ for the nonlinearity parameter. This parameter must not be confused with the magnitude of the space position $\brm{r}$. The nonlinearity parameter always appears as a power or a subscript, respectively, and when read in its context, the meaning of $r$ should be clear. The main benefit of the nonlinear RRN is that it allows to map out the important fractal substructures of percolation clusters by taking the appropriate limits with respect to $r$. Of course, one retrieves the linear RRN for $r\to 1$. We will refer to the model introduced in Ref.~\cite{harris_87} as the nonlinear HL model.

\subsubsection{Hamiltonian}
To capture the RRN, one distributes the $n+1=(2M+1)^{D}$ states of the $(n+1)$-state Potts model on a $D$-dimensional periodic lattice (torus) with site-coordinates $\vec{\theta}=\Delta\theta(j_{1},\ldots,j_{D})$, where $j_{i}=-M,-M+1,\ldots,M$. The combination
\begin{align}
\sqrt{M}\Delta\theta :=\theta_{0}
\end{align}
is a free parameter of the theory which plays a subtle role. We will comment on this role as we move along. The order parameter field of the (nonlinear) HL model is given by the generalized Potts spin $\varphi_{\vec{\theta}}$ subject to the constraint
\begin{align}
\label{constraint}
\sum_{\vec{\theta}}\varphi_{\vec{\theta}}=0\, .
\end{align}
It is related to its Fourier transformed counterpart via
\begin{align}
\label{voltageCurrentTrafo}
\varphi_{\vec{\theta}}  &  =\frac{1}{(2M+1)^{D/2}}\left.  \sum_{\vec{\lambda}%
}\right.^\prime  \mathrm{e}^{i\vec{\theta}\cdot\vec{\lambda}}\tilde{\varphi
}_{\vec{\lambda}}\,,
\\
\tilde{\varphi}_{\vec{\lambda}}  &  =\frac{1}{(2M+1)^{D/2}}\sum_{\vec{\theta}%
}\mathrm{e}^{-i\vec{\theta}\cdot\vec{\lambda}}\varphi_{\vec{\theta}}\,,
\end{align}
where $\vec{\lambda}=(\lambda_{1},\ldots\lambda_{D})$ with
\begin{equation}
\label{defLambdaAlpha}
\lambda_{\alpha}=\frac{2\pi l_{\alpha}}{(2M+1)\Delta\theta}\operatorname{mod}2\pi/\Delta
\theta \, ,
\end{equation}
and $l_{\alpha}=-M,-M+1,\ldots,M$. The prime in Eq.~(\ref{voltageCurrentTrafo}) indicates that $\vec{\lambda} = \vec{0}$ is excluded from the sum. In the following, we will drop this prime for notational simplicity unless stated otherwise. Through this Fourier transformation, the constraint (\ref{constraint}) translates into $\tilde{\varphi}_{\vec{\lambda}=0}=0$. Expressed in terms of $\tilde{\varphi}_{\vec{\lambda}}$, the nonlinear HL model takes on the form
\begin{align}
\mathcal{H}&=\int d^{d}r  \Bigg\{  \frac{1}{2}\sum_{\vec{\lambda}
} \tilde{\varphi}_{{-\vec{\lambda}}}\left[  \tau
+w\Lambda_r(\vec{\lambda})-\nabla^{2}\right]  \tilde{\varphi}_{{\vec{\lambda}}
}
\nonumber \\
&\qquad\qquad +\frac{g}{6}\left.  \sum_{\vec{\lambda},\vec{\lambda}}\right.
\tilde{\varphi}_{{\vec{\lambda}}}\tilde{\varphi}_{\vec{\lambda
}^{\prime}}\tilde{\varphi}_{-\vec{\lambda}-\vec{\lambda}^{\prime}}\Bigg\} ,
\end{align}
where
\begin{align}
\Lambda_r(\vec{\lambda})=-\sum_{\alpha=1}^{D}(-\lambda_{\alpha}
^{2})^{(r+1)/2} .
\end{align}
The control parameter $w$ is proportional to the resistance of the occupied bonds. The parameter $r$ specifies the non-linearity of the network with $r=1$ corresponding to the linear RRN. The HL model constitutes a Potts model with a specific quadratic symmetry breaking. Purely geometrical percolation is retrieved for $w\to 0$ provided one takes the limit $D\to 0$ with $M$ finite. In contrast, for the model to describe the (nonlinear) RRN, one has to take the limits $D \to 0$ and $M\to \infty$ (keeping $\theta_0$ fixed) with the order of the limits being of crucial importance. In the following, we will refer to this double limit as the RRN-limit.

There are some subtle, however important, features of the theory associated with the RRN-limit. As long as $M$ is finite, we have a model for percolation with quadratic symmetry breaking which can be handled by simple insertions of the operator $w\sum_{\vec{\lambda}} \tilde{\varphi}_{-\vec{\lambda}}\Lambda_r(\vec{\lambda})\tilde{\varphi}_{\vec{\lambda}}$. However, in the continuum-limit of the $\theta$-variables,  $M\to\infty$, sums of the form $\sum_{\vec{\lambda}}\lambda_{\alpha_1}...\lambda_{\alpha_k}$ diverge like $M^{k/2}$ even if $D\to 0$, and a more careful treatment is in order \cite{harris_lubensky_84,harris_lubensky_87b}. In the RRN-limit, the free parameter $\theta_{0}$ becomes obsolete, or in the language of field theory, it becomes a redundant parameter.
As such, it is possible to eliminate it via rescaling of $w$. That this elimination can indeed be achieved hinges on the fact that in perturbation theory in conjunction with the RRN-limit, $w$ and $\Lambda_r(\vec{\lambda})$ always appear in the combination $w \Lambda_r(\vec{\lambda})$. This in turn implies that only this combination, as opposed to $w$ or  $\Lambda_r(\vec{\lambda})$ alone, has a well defined engineering dimension, namely $2$ like $\tau$ or $\nabla^2$.

\subsubsection{Connection probability and average nonlinear resistance}
\label{sec:ConProbAndAvNonlRes}
The success of the HL model and its nonlinear generalization stems, at least in part, from the fact that the two-point correlation function
\begin{eqnarray}
\label{defCorr}
G(\brm{r}, \vec{\lambda}, \tau, w  ) = \left\langle \tilde{\varphi}_{-{\vec{\lambda}}} ( \brm{r}) \tilde{\varphi}_{{\vec{\lambda}}} (\brm{0}) \right\rangle
\end{eqnarray}
provides for an elegant and efficient route to calculate the average resistance between two points on the same cluster. In the following, we will use the notation $G( \brm{r}, \tau, w \Lambda_r(\vec{\lambda})  )$ for the two-point function, as we can because $\vec{\lambda}$ and $w$ always appear in the combination $w \Lambda_r(\vec{\lambda})$, see above. That the two-point function~(\ref{defCorr}) is helpful in calculating the average resistance becomes transparent if one notices that it constitutes a generating function:
\begin{align}
\label{genFkt}
G(\brm{r}, \tau, w \Lambda_r(\vec{\lambda})  )
&=\left[ \chi (\brm{r} |C ) \, \exp \big(-w \Lambda_r (\vec{\lambda}) {R}_r (\brm{r} | C ) \big)\right]_{av}
\nonumber \\
&= \chi(\brm{r} )\left[{\exp }\big( -w \Lambda_r (\vec{\lambda}){R}_r (\brm{r} | C )\big)\right]_{av}^\prime \, .
\end{align}
where $R_r (\brm{r} | C )$ is proportional to the total nonlinear resistance between an arbitrary point ${\bf r}$ and $\brm{0}$ for a given random percolation configuration $C$.
Here, we have absorbed the resistance $\rho$ of the conducting microscopic bonds into the control parameter $w$ such that $R_r (\brm{r} | C ) \propto \rho^0$. $\chi (\brm{r} |C )$ is an indicator function which is one if ${\mathbf{r}}$ is connected to $\brm{0}$ in configuration $C$, and zero otherwise. $\left[\cdots\right]_{av}$ denotes the disorder average over all configurations of the diluted lattice. $\left[\cdots\right]_{av}^\prime$ stands for disorder averaging conditional to the constraint that ${\mathbf{r}}$ and $\brm{0}$ are connected.
\begin{align}
\chi (\brm{r}) =\left[\chi (\brm{r} |C )\right]_{av}
\end{align}
is the usual percolation correlation function, i.e., the probability for any two sites a distance $|{\mathbf{r}}|$ apart being connected. From Eq.~(\ref{genFkt}) it follows that one can extract the average resistance
\begin{eqnarray}
M_{r}(\brm{r} )  = \left[R_r (\brm{r} | C )\right]_{av}^\prime
\end{eqnarray}
and its moments simply via taking the derivatives of $G( \brm{r}, \tau, w \Lambda_r(\vec{\lambda}))$ with respect to $w\Lambda_r (\vec{\lambda})$.

As mentioned above, the great value of the nonlinear RRN lies in the fact that it can be used to map out different fractal substructures of percolation clusters. Inspection of the overall dissipated electric power shows readily that 
\begin{eqnarray}
\label{relDB}
\lim_{r \to -1^+} M_{r}(\brm{r})  \sim M_{bb}(\brm{r}) \, ,
\end{eqnarray}
where $M_{bb}$ is proportional to the average mass (number of bonds) of the backbone (between any two site a distance $|\brm{r}|$ apart, of course). Moreover, it has been shown by Blumenfeld and Aharony~\cite{blumenfeld_aharony_85} that
\begin{eqnarray}
\lim_{r \to \infty} M_{r}(\brm{r})  \sim M_{\text{red}}(\brm{r}) \, ,
\end{eqnarray}
where $M_{\text{red}}$ stands for the mass of the red bonds and
\begin{eqnarray}
\label{relMin}
\lim_{r \to 0^+} M_{r}(\brm{r})  \sim M_{\text{min}}(\brm{r}) \, ,
\end{eqnarray}
where $M_{\text{min}}$ is the mass (length) the shortest self-avoiding walk (SAW). Likewise, on can show that
\begin{eqnarray}
\label{relMax}
\lim_{r \to 0^-} M_{r}(\brm{r})  \sim M_{\text{max}}(\brm{r}) \, ,
\end{eqnarray}
where $M_{\text{max}}$ is the mass (length) of the longest SAW. As alluded to above, it is not known how to extract the mass (length) $M_{\text{SAW}}$ of the average SAW from $M_r$. However, we recently proposed~\cite{janssen_stenull_SAW_2007} a method for calculating $M_{\text{SAW}}$ based on the RRN that does not invoke $M_r$ directly. The idea behind this method is exploit the real-world interpretation of the RRN Feynman diagrams, which will be explained in the next subsection, by placing SAWs onto these diagrams (for details see Ref.~\cite{janssen_stenull_SAW_2007}). Using this approach, one can calculate $M_{\text{SAW}}$ in much the same way as $M_{\text{min}}$ and $M_{\text{max}}$.

\subsubsection{Renormalization, real world interpretation of Feynman diagrams and scaling}
\label{renScalingInt}
From the above, it should be clear that the renormalization factors $Z$, $Z_\tau$ and $Z_u$ for the RRN are identical to those for the Potts model with  $n+1=(2M+1)^{D}$ states. The control parameter $w$, however, needs an independent renormalization
\begin{align}
w\to\mathring{w}= Z^{-1}Z_{w} w
\end{align}
with the renormalization factor $Z_w$ having no counterpart in the usual Potts model. This renormalization can be calculated in an elegant and efficient way by using our real
world interpretation of the HL model's Feynman diagrams~\cite{stenull_janssen_oerding_99,janssen_stenull_oerding_99,janssen_stenull_99,stenull_2000}. This interpretation is based on the observation that the diagrammatic perturbation theory of the HL model in the RRN-limit can be formulated in such a way that the Feynman diagrams resemble real RRNs. In this approach, the diagrams feature conducting propagators
corresponding to occupied, conducting bonds and insulating propagators
corresponding to open bonds. The conducting bonds carry replica currents $- i \vec{\lambda}$. The resistance of a conducting bond is given by its Schwinger parameter in a parametric representation of the propagators. Because of the interpretation of $\vec{\lambda}$ (apart from the factor $-i$) as a replicated current, we refer to the space in which $\vec{\lambda}$ takes on its values as the replicated current space. Likewise, we refer to the space in which the conjugate variable $\vec{\theta}$ lives as the replicated voltage space.

When this real-world interpretation is used, the task of calculating $Z_w$ is by and large reduced to calculating the total nonlinear resistance of the conducting Feynman diagrams. To one-loop order, one obtains \cite{harris_87,janssen_stenull_oerding_99}
\begin{align}
Z_{w}  &  =1+\left(  1-\int_{0}^{1}dx\,c_{r}(x)\right)  \frac{u}{\varepsilon
} \, .
\end{align}
The function $c_{r}(x)$ is given by
\begin{align}
\label{defCrGreaterZero}
c_{r}(x)  &  =\frac{x\left(  1-x\right)  }{\left[  x^{1/r}+\left(  1-x\right)
^{1/r}\right]  ^{r}}\,
\end{align}
Note that it vanishes for $x=0$ und $x=1$ if $r$ is positive. For negative $r$, $c_{r}(x)$ it takes the form
\begin{equation}
\label{defCrSmallerZero}
c_{r}(x)=\left[  x^{1/\left\vert r\right\vert }+\left(  1-x\right)
^{1/\left\vert r\right\vert }\right]  ^{\left\vert r\right\vert }\,,
\end{equation}
which is equal to $1$ for $x=0$ und $x=1$. Here, we encounter another manifestation of the  discontinuity at $r=0$ which, physically, is related to the different types of SAWs on the percolation clusters.

Following standard procedures of renormalized field theory, one can extract from the above renormalizations a Gell-Mann--Low renormalization group equation (RGE) that governs the scaling behavior of the correlation functions. For the 2-point functions, in particular, this RGE leads to the well known result 
\begin{equation}
\label{ScaleForm2Point}
G(\mathbf{r},\tau,w\Lambda_r(\vec{\lambda})) = \left\vert
\mathbf{r}\right\vert ^{2-d-\eta}\hat{G}\big(\tau\left\vert \mathbf{r}\right\vert
^{1/\nu},w\Lambda_r(\vec{\lambda})\left\vert \mathbf{r}\right\vert
^{\phi_r/\nu}\big)\,,
\end{equation}
were $\hat{G}$ is, up to scaling factors, a universal scaling function. The critical exponents $\eta$ and $\nu$ are known to 3-loop order~\cite{alcantara_80}. To 2-loop order, they are given by
\begin{subequations}
\begin{align}
\eta &= -\frac{1}{21}\varepsilon - \frac{206}{9261}\varepsilon^2 ,
\\
\nu &  = \frac{1}{2} + \frac{5}{84}\varepsilon + \frac{589}{37044}\varepsilon^2 .
\end{align}
\end{subequations}
The resistance exponent $\phi_r$ is know for general $r$ to 1-loop order up to a remaining integral~\cite{harris_87},
\begin{align}
\phi_r = 1 + \frac{\varepsilon}{14} \, \int_{0}^{1}dx\,c_{r}(x) \, ,
\end{align}
with $c_{r}(x)$ given by Eq.~(\ref{defCrGreaterZero}). For specific values of $r$, $\phi_r$ is known to higher order. For the linear RRN, the resistance exponent $\phi = \phi_1$ is known to 2-loop order~\cite{lubensky_wang_1985,stenull_janssen_oerding_99,stenull_2000}. In the limit $r\to -1^+$ relevant for the backbone, we calculated $\phi_r$ to 3-loop order~\cite{janssen_stenull_oerding_99,janssen_stenull_99,stenull_2000}. In the limit $r\to \infty$ relevant for the red bonds, we showed explicitly to 3-loop order that $\phi_r = 1$~\cite{janssen_stenull_oerding_99,janssen_stenull_99,stenull_2000} as had to be expected from rigorous results by Coniglio~\cite{coniglio_81_82}. Furthermore, we calculated $\phi_r$ in the limit $r\to 0^+$ describing the shortest SAW~\cite{janssen_85,janssen_stenull_oerding_99,janssen_stenull_99,stenull_2000} as well as the exponents governing the longest and the average SAW to 2-loop order~\cite{janssen_stenull_SAW_2007}.

\subsubsection{Probability distribution for the nonlinear resistance -- definition and mean field theory}
\label{probDistDefAndMF}
As discussed in the introduction, it is useful for a comprehensive characterization of the properties of the RRN and the various fractal substructures of percolation clusters to go beyond just studying the average nonlinear resistance or its higher moments and to address the more general problem of the form of the distribution function of the nonlinear resistance.

The distribution $P_r(R,\brm{r})$ of the conditional probability that the total nonlinear resistance between $\brm{r}$ and $\brm{0}$ is ${R}_r$ given that $\brm{r}$ and $\brm{0}$ are on the same cluster can be defined as
\begin{equation}
\label{defResDistr}
P_r(R,\brm{r})=\left[  \delta(R-{R}_{r}(\brm{r}|C) \right]_{av}^\prime \, .
\end{equation}
Comparison with Eq.~(\ref{genFkt}) reveals that
\begin{equation}
\label{defPi}
\Pi(z,\brm{r}):=\frac{G(\brm{r},\tau,z)}{G(\brm{r},\tau,0)}
=\int_{0}^{\infty}dR\,\mathrm{e}^{-zR}P_r(R,\brm{r})
\end{equation}
is the Laplace transform of $P_r(R,\brm{r})$. Thus, one can calculate $P_r(R,\brm{r})$ via calculating the two point correlation function $G( \brm{r}, \tau, w \Lambda_r(\vec{\lambda})=:z)$ and the applying inverse Laplace transformation to $\Pi(z,\brm{r})$,
\begin{align}
\label{inverseTrafoYieldingP}
P_r(R,\mathbf{r}) &  =\frac{1}{2\pi i}\int_{\sigma-i\infty}^{\sigma+i\infty
}dz\,\mathrm{e}^{zR}\, \Pi(z,\brm{r})\,.
\end{align}
This is the main route that we will follow in the remainder of this paper.

The distributions of the various fractal masses can be defined by modifying Eq.~(\ref{defResDistr}) in an evident manner. These distributions can be extracted from $P_r(R,\brm{r})$ simply by taking the appropriate limit with respect to $r$ except for the distribution $P_{\text{SAW}} (L, \brm{r})$ of the length $L$ of the average SAW. Nevertheless, by combining the route described above with the method proposed in Ref.~\cite{janssen_stenull_SAW_2007}, we can calculate $P_{\text{SAW}} (L, \brm{r})$ in much the same way as the distributions of the other fractal masses.

Before embarking on field theory, we conclude this review section by considering mean field theory. At 0-loop order, the 2-point correlation function is given simply by the Fourier transform of the Gaussian propagator,
\begin{align}
\label{Gaussprop}
G(\mathbf{r},\tau,z)&=\int_{\mathbf{q}}\frac{\exp\left(  i\mathbf{q\cdot
r}\right)  }{q^{2}+\tau+z} =\frac{1}{(2\pi)^{d/2}}\left(  \frac{\tau+z}{\brm{r}^{2}
}\right)  ^{(d-2)/4}
\nonumber \\
&\times K_{(d-2)/2}(|\brm{r}|\sqrt{\tau+z})\,,
\end{align}
where $\int_{\mathbf{q}}$ is an abbreviation for $\frac{1}{(2\pi)^{d}} \int d^d q$, and where $K_{\alpha}(x)$ is the modified Bessel function (Basset function) with index $\alpha$. Using this 2-point function in conjunction with Eq.~(\ref{defPi}) in Eq.~(\ref{inverseTrafoYieldingP}) and introducing the quantity
\begin{equation}
C_{\alpha}(x)=\frac{\Gamma(\alpha)}{2}\left(  \frac{2}{x}\right)  ^{\alpha
}K_{\alpha}(x)^{-1}\,
\end{equation}
which becomes unity at criticality, $C_{\alpha}(0)=1$, we obtain~\cite{footnote:meanFieldDist}
\begin{align}
\label{eq:meanFieldDist}
P_r(R,\mathbf{r}) &  =\frac{C_{(d-2)/2}(|\brm{r}|\sqrt{\tau})}{\Gamma((d-2)/2)} \,
R^{-1} \, \frac{\exp\left(  -\tau R-\brm{r}^{2}/(4R)\right)  }{\left(  4R/\brm{r}^{2}\right)
^{(d-2)/2}}
\end{align}
for the resistance distribution in mean field theory. Note that, from this expression, one can read off the mean field results for the correlation length and resistance exponents, $\nu = 1/2$ and $\phi_r = 1$. The mean-field result for $P_{\text{SAW}} (L, \brm{r})$ is of the same form as Eq.~(\ref{eq:meanFieldDist}) with $R$ replaced by $L$. This implies the well known mean-field result $\nu_{\text{SAW}} =1/2$.

\begin{figure}
\centerline{\includegraphics[width=7.8cm]{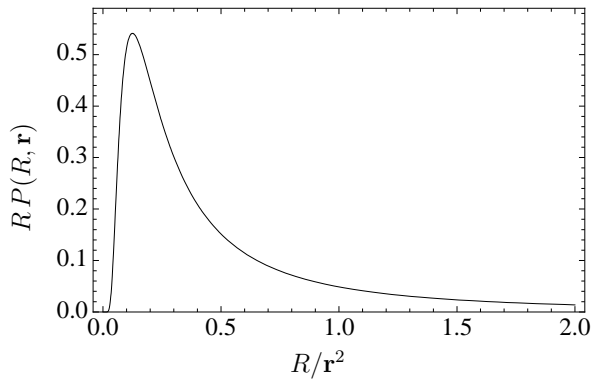}}
\caption{Form of the resistance distribution function at criticality as given in mean field theory by Eq.~(\ref{eq:meanFieldDist}) with $\tau = 0$ and $d=6$.}
\label{fig:meanFieldDist}
\end{figure}

At criticality, the distribution function takes on the scaling form~(\ref{scaleFromPrCrit}). The corresponding mean field scaling function is plotted in Fig.~\ref{fig:meanFieldDist}. For large and small arguments, it behaves asymptotically as the leading term in Eq.~(\ref{OmegaTsmallS}) and Eq.~(\ref{OmegaTlargeS}), respectively, with $\nu/(\phi_r-\nu) =1$ and $d\nu-1 = (d-2+2\eta)\nu/[2(\phi_r-\nu)]= 2$.

As mentioned above, HL based their calculation on the assumption that the unconditional distribution function $P_{\text{uc}}(R,\mathbf{r}) = \chi(\brm{r}) P(R,\mathbf{r})$ becomes independent of $\brm{r}$ for $s =R/\brm{r}^2 \gg 1$. Noting from Eq.~(\ref{ScaleForm2Point}) that $\chi(\brm{r}) \sim |\brm{r}|^{-d+2-\eta}$ and recalling from Eq.~(\ref{OmegaTsmallS}) that $\Omega_r(s) \sim s^{- g_1} +\cdots $ with $g_1 = (d\nu -1)/\phi_r$ in this limit, we find $P_{\text{uc}}(R,\mathbf{r})  \sim |\brm{r}|^{g^\prime}$ with $g^\prime = 2 -\eta - 1/\nu$. In mean field theory, where $\eta = 0$ and $\nu =1/2$, $g^\prime$ vanishes so that $P_{\text{uc}}$ is indeed independent of $\brm{r}$. Beyond mean field theory, however, $g^\prime$ is non-zero, and the assumption by HL is incorrect. Because of this shortcoming, the HL result $g_1^{\text{HL}}= (d -2 +\eta)\nu/\phi_1$ for $g_1$ is not correct for $d<6$, although it produces the correct mean-field value.

\section{Representations, composite operators and their renormalization}
\label{representations}
Now we turn to renormalized field theory to study the connection probability and the nonlinear resistance distribution beyond mean field theory. In particular, we will calculate their asymptotic forms in the limits $x= \tau| \brm{r}|^{1/\nu} \ll 1$ and $s = R/|\brm{r}|^{\phi_r/\nu} \gg 1$, respectively. These calculations are quite non-trivial because we need to determine the short-distance behavior of the 2-point correlation function. This behavior can be analyzed systematically with help of the well established short-distance expansion~\cite{amit_zinn-justin}. We will explain the application of this technique to the current problem in some detail in Sec.~\ref{shortDistanceExpansion}. It turns out, that the composite field or composite operator $\frac{1}{2} \tilde{\varphi}_{\vec{\lambda}} ( \brm{r}) \tilde{\varphi}_{{-}\vec{\lambda}} (\brm{r})$ plays a pivotal role in the short-distance expansion. In this section we discuss the renormalization of this and related operators employing group theory \cite{Ham62}. Our approach is guided by the work of Wallace and Young \cite{WaYo78}, and we review parts of their work in a language that is adapted to the problem at hand.

We are interested in the direct product of the order parameter with itself. As discussed below Eq.~(\ref{PottsConstraintOriginal}), the order parameter  $\varphi_{\vec{\theta}}$ is an object which transforms as the fundamental irreducible representation $(n,1)$ of the symmetric group $S_{n+1}$, if the invariant identity representation $(n+1)$ is ruled out through the constraint $\sum_{\vec{\theta}}\varphi_{\vec{\theta}}=0$. The product of the fundamental representation with itself is reducible and it can be decomposed into a sum of irreducible representations,
\begin{equation}
(n,1)\times(n,1)=(n+1)+(n,1)+(n-2,2)+(n-1,1^{2}),
\end{equation}
comprising the identity representation $(n+1)$, the fundamental representation $(n,1)$,  the symmetric tensor representation $(n-2,2)$ and antisymmetric tensor representation $(n-1,1^{2})$. The latter is absent for a product of two identical fields. Applied to the product $\varphi_{\vec{\theta}}\varphi_{\vec{\theta^{\prime}}}$, the decomposition leads to the identity representation
\begin{equation}
\mathcal{B}=\frac{1}{2}\sum_{\vec{\theta}}\varphi_{\vec{\theta}}^{2}\,,
\end{equation}
the fundamental representation
\begin{equation}
\Psi_{\vec{\theta}}=\frac{\sqrt{n+1}}{2}\varphi_{\vec{\theta}}^{2}-\frac{\mathcal{B}}{\sqrt{n+1}}\,,
\end{equation}
and the symmetric tensor representation
\begin{equation}
\Phi_{\vec{\theta},\vec{\theta}^\prime}=(1-\delta_{\vec{\theta},\vec{\theta}^\prime})\bigg\{
\frac{1}{2}\varphi_{\vec{\theta}}\varphi_{\vec{\theta}^\prime}
+\frac{\Psi_{\vec{\theta}}+\Psi_{\vec{\theta}^\prime}}{(n-1)\sqrt{n+1}}
+\frac{\mathcal{B}}{n(n+1)}\biggr\}\,.
\end{equation}

To simplify the discussion to follow, it is useful to switch from replicated voltage space to replicated current space via the Fourier transformation~(\ref{voltageCurrentTrafo}).  Then, the above operators take on the form
\begin{subequations}
\label{FourierTransIrredOps}
\begin{align}
\mathcal{B} &= \frac{1}{2} \sum_{\vec{\kappa}} \tilde{\varphi}_{-\vec{\kappa}}  \tilde{\varphi}_{\vec{\kappa}} \, ,
\\
\tilde{\Psi}_{\vec{\lambda}} &= \frac{1}{2} \sum_{\vec{\kappa}} \tilde{\varphi}_{\vec{\lambda}-\vec{\kappa}}  \tilde{\varphi}_{\vec{\kappa}}  \, ,
\\
\label{FourierTransIrredOps3}
\tilde{\Phi}_{\vec{\lambda}, \vec{\lambda^{\prime}}} &=\frac{1}{2} \tilde{\varphi}_{\vec{\lambda}}  \tilde{\varphi}_{\vec{\lambda^{\prime}}}
 - \frac{\tilde{\Psi}_{\vec{\lambda}+\vec{\lambda^{\prime}}}}{n-1} \,
-  \frac{\mathcal{B}}{n} \, \delta_{\vec{\lambda}+\vec{\lambda}^{\prime},\vec{0}} \, ,
\end{align}
\end{subequations}
if $\vec{\lambda}$ and $\vec{\lambda}^{\prime}$ are non-zero. Otherwise
$\tilde{\Psi}_{\vec{0}}=\tilde{\Phi}_{\vec{\lambda},\vec{0}}=\tilde{\Phi}_{\vec{0},\vec{\lambda}}=0$. Hence, there does not exist any component with zero current.

As mentioned above, it will be the operator
\begin{align}
\mathcal{A}_{\vec{\lambda}}=: \frac{1}{2} \tilde{\varphi}_{\vec{\lambda}} ( \brm{r}) \tilde{\varphi}_{-\vec{\lambda}} (\brm{r})
\end{align}
which plays a central role in the SDE. Evidently, this operator satisfies $\mathcal{A}_{\vec{\lambda}=0}=0$. As we will discuss in the following, the renormalization of $\mathcal{A}_{\vec{\lambda}}$ is intimately related to that of the operators $\mathcal{B}$ and $\tilde{\Phi}_{\vec{\lambda}, \vec{\lambda^{\prime}}}$.
Equation~(\ref{FourierTransIrredOps}) implies that the operators $\mathcal{A}_{\vec{\lambda}}$, $\mathcal{B}$ and $\tilde{\Phi}_{\vec{\lambda}, -\vec{\lambda}}$ are related for $\vec{\lambda}\neq 0$ via
\begin{align}
\label{relABPhi}
\mathcal{A}_{\vec{\lambda}} = \frac{1}{2} \,  \tilde{\Phi}_{\vec{\lambda}, -\vec{\lambda}} + \frac{1}{n} \, \mathcal{B} \, .
\end{align}
Note that the decomposition of $\mathcal{A}_{\vec{\lambda}}$ in irreducible parts contains a factor $1/n$ which could be potentially problematic in the limit $n\to 0$. This factor, however, cancels in perturbation theory.

Form the group theoretic considerations above, we draw the three important conclusions. (i) The operator $\tilde{\Psi}_{\vec{\lambda}}$ can mix additively with the order parameter field $\tilde{\varphi}_{\vec{\lambda}}$ and its derivatives because both transform according to the same irreducible representation $(n,1)$. (ii) The operators $\mathcal{B}$ and $\tilde{\Phi}_{\vec{\lambda}, \vec{\lambda^{\prime}}}$ are renormalizable individually by multiplicative renormalization, because they belong to representations which are different from one another and $(n,1)$. (iii) The operator $\mathcal{A}_{\vec{\lambda}}$,
which will play a central role in the SDE, contains both the independent renormalizations of $\mathcal{B}$ and $\tilde{\Phi}_{\vec{\lambda}, -\vec{\lambda}}$ and mixes these.

Equipped with this information, we will now determine the renormalizations of the various composite operators. With the renormalization scheme~(\ref{reno}), the renormalized version of the generalized Potts Hamiltonian is given by
\begin{align}
\mathcal{H}&=\int d^{d}r\,\Bigg\{  \frac{1}{2} \sum_{\vec{\lambda}}
\tilde{\varphi
}_{_{-\vec{\lambda}}}\left[  Z_{\tau}\tau-Z\nabla^{2}\right]
\tilde{\varphi
}_{_{\vec{\lambda}}}
\nonumber\\
&+Z_{u}^{1/2}\, \frac{g}{6}\sum_{\vec{\lambda}%
,\vec{\lambda}^{\prime}}\tilde{\varphi
}_{_{\vec{\lambda}}}\tilde{\varphi}
_{\vec{\lambda}^{\prime}}\tilde{\varphi
}_{-\vec{\lambda}-\vec{\lambda}^{\prime}}\Bigg\}  \,.
\end{align}
From this renormalized Hamiltonian, we can infer without much effort the renormalizations of $\mathcal{B}$ and $\tilde{\Psi}_{\vec{\lambda}}$. As far as $\mathcal{B}$  is concerned, we note that if $\tau$ is viewed as a function of $\brm{r}$ one can generate insertions of the operator $-Z_\tau \mathcal{B}$ into renormalized correlation functions via functional differentiation with respect to $\tau(\brm{r})$. Thus, the renormalized version $\mathcal{B}_R$ of $\mathcal{B}$ is given by
\begin{equation}
\mathcal{B}_R=Z_{\tau}\, \mathcal{B}=Z_{\tau}Z^{-1} \, \frac{1}{2}  \sum_{\vec{\kappa}}
\mathring{\tilde{\varphi}}_{-\vec{\kappa}}\mathring{\tilde{\varphi}}_{\vec{\kappa}}\,.
\end{equation}
Turning to $\tilde{\Psi}_{\vec{\lambda}}$, we note that the functional derivative with respect to $\tilde{\varphi}_{\vec{\lambda}} (\brm{r})$ leads to the equation of motion
 \begin{align}
&\left\langle\frac{\delta\mathcal{H}}{\delta\tilde{\varphi}_{_{-\vec{\lambda}}}(\mathbf{r}%
)}\tilde{\varphi}(\mathbf{r}_{1})\cdots\tilde{\varphi}(\mathbf{r}_{N})\right\rangle=\sum_{i=1}%
^{N}\langle\tilde{\varphi}(\mathbf{r}_{1})\cdots\tilde{\varphi}(\mathbf{r}_{i-1})
\nonumber \\
&\times \,\tilde{\varphi}(\mathbf{r}_{i+1})\cdots\tilde{\varphi}(\mathbf{r}_{N})\rangle \, \delta
(\mathbf{r-r}_{i})\,,
\end{align}
where we have suppressed the current-index of most of the fields for notational simplicity. As a consequence of this equation of motion, insertions of the operator
\begin{equation}
\frac{\delta\mathcal{H}}{\delta\tilde{\varphi}_{_{-\vec{\lambda}}}}=Z_{\tau}\tau
\tilde{\varphi}_{_{\vec{\lambda}}}-Z\nabla^{2}\tilde{\varphi}_{_{\vec{\lambda}}}+Z_{u}^{1/2}\frac{g}%
{2}\sum_{\vec{\kappa}}\tilde{\varphi}_{\vec{\lambda}-\vec{\kappa}}%
\varphi_{\vec{\kappa}}\,,
\end{equation}
into renormalized correlation functions are already completely renormalized. This implies that the renormalized version $\tilde{\Psi}_{\vec{\lambda}; R}$ of $\tilde{\Psi}_{\vec{\lambda}}$ is given by
\begin{align}
\tilde{\Psi}_{\vec{\lambda};R}  &  =Z_{u}^{1/2}\tilde{\Psi}_{\vec{\lambda}}+\frac{1}{g}\left(  Z_{\tau
}-1\right)  \tau\, \tilde{\varphi}_{_{\vec{\lambda}}}-\frac{1}{g}\left(  Z-1\right)  \nabla
^{2}\tilde{\varphi}_{_{\vec{\lambda}}}
\nonumber\\
&  =Z_{u}^{1/2}Z^{-1}\sum_{\vec{\kappa}}\mathring{\tilde{\varphi}
}_{\vec{\lambda}-\vec{\kappa}}\mathring{\tilde{\varphi}}_{\vec{\kappa}}+\frac{1}{g}Z^{-1/2}\left(
Z_{\tau}-1\right)  \tau\, \mathring{\tilde{\varphi}}_{_{\vec{\lambda}}}
\nonumber\\
&-Z^{-1/2}\frac{1}{g}\left(  Z-1\right)  \nabla^{2}\mathring{\tilde{\varphi}}_{_{\vec{\lambda}}}\,.
\end{align}
As noted above, $\tilde{\Psi}_{\vec{\lambda}}$ mixes under renormalization additively with  $\tilde{\varphi}_{\vec{\lambda}}$ and its derivatives.

Next, we move on to $\mathcal{A}_{\vec{\lambda}}$ and $\tilde{\Phi}_{\vec{\lambda}, -\vec{\lambda}}$. Our strategy is to determine the renormalization of $\mathcal{A}_{\vec{\lambda}}$ with the help of that of $\tilde{\Phi}_{\vec{\lambda}, -\vec{\lambda}}$ and $\mathcal{B}$ via relation~(\ref{relABPhi}), and we will switch back and forth between considering $\mathcal{A}_{\vec{\lambda}}$ and $\tilde{\Phi}_{\vec{\lambda}, -\vec{\lambda}}$ to achieve this goal. As we know from our group theoretic considerations above, the renormalization of $\tilde{\Phi}_{\vec{\lambda}, -\vec{\lambda}}$ is independent of that of $\mathcal{B}$ and $\tilde{\Psi}_{\vec{\lambda}}$, and thus it is renormalizable by multiplicative renormalization. Also because of this independence, we cannot simply extract the renormalization of $\tilde{\Phi}_{\vec{\lambda}, -\vec{\lambda}}$ from the above. As the typical quadratic symmetry-breaking term in the Hamiltonian, the tensor $\tilde{\Phi}_{\vec{\lambda}, \vec{\lambda}^{\prime}}$ determines the so-called crossover exponent. Thus, we denote its renormalization factor by $Z_c$ so that
\begin{equation}
\tilde{\Phi}_{\vec{\lambda},\vec{\lambda}^{\prime};R}=Z_{c}\tilde{\Phi}_{\vec{\lambda},\vec{\lambda}^{\prime}}\,.
\end{equation}

To approach the renormalization of $\mathcal{A}_{\vec{\lambda}}$ through perturbation theory, we consider its insertions into vertex functions $\Gamma_{\vec{\kappa},-\vec{\kappa}}^{(2)}$, i.e., $\Gamma_{\vec{\kappa}
,-\vec{\kappa};\mathcal{A}_{\vec{\lambda}}}^{(2)}$. According to the above, $\mathcal{A}_{\vec{\lambda}}$ itself is not renormalizable multiplicatively as it requires additive contributions from $\mathcal{B}$, i.e.,
the renormalized version of $\mathcal{A}_{\vec{\lambda}}$ will be of the type
\begin{equation}
\label{formRenA}
\mathcal{A}_{\vec{\lambda};R}=Z^{\prime}\mathcal{A}_{\vec{\lambda}}+Y\mathcal{B}
\end{equation}
Any result for $\Gamma_{\vec{\kappa} ,-\vec{\kappa};\mathcal{A}_{\vec{\lambda}}}^{(2)}$ produced by diagrammatic perturbation theory has to be of the form
\begin{equation}
\label{formGamma2Ains}
\Gamma_{\vec{\kappa},-\vec{\kappa};\mathcal{A}_{\vec{\lambda}}}^{(2)}=\frac{1}{2}\left(
\delta_{\vec{\lambda},\vec{\kappa}}+\delta_{\vec{\lambda},-\vec{\kappa}}\right)  \left(
1+A(n)\right)  +B(n)\,,
\end{equation}
where $A(n)$ and $B(n)$ are polynomials in $n$ which at each order in $n$ have a perturbation expansion in powers of $u$ containing $\varepsilon$-poles. Summing both sides of Eq.~(\ref{formGamma2Ains}) over $\vec{\lambda}$ produces
\begin{equation}
\Gamma_{\vec{\kappa},-\vec{\kappa};\mathcal{B}}^{(2)}=\left(  1+A(n)\right)  +nB(n)\,.
\end{equation}
We already know that $\Gamma_{\vec{\kappa},-\vec{\kappa};\mathcal{B}}^{(2)}$ is renormalized by multiplication with $Z_{\tau}$, i.e., $Z_{\tau}\Gamma_{\vec{\kappa},-\vec{\kappa};\mathcal{B}}^{(2)}=\Gamma_{\vec{\kappa},-\vec{\kappa};\mathcal{B}_{R}}^{(2)}$ is free of $\varepsilon$-poles, and thus
\begin{equation}
\label{finite1}
\Gamma_{\vec{\kappa},-\vec{\kappa} ;\mathcal{B}_{R}}^{(2)} = Z_{\tau}+Z_{\tau}\left(  A(n)+nB(n)\right)  = \mbox{finite} \,,
\end{equation}
from which $Z_{\tau}$ can be constructed order for order in $u$ with polynomial factors in $n$.

Equipped with this information, we now insert the renormalized version of $\mathcal{A}_{\vec{\lambda}}$ as given in Eq.~(\ref{formRenA}) into the renormalized two-point vertex function and determine $Z^{\prime}$ and $Y$ such that $\Gamma_{\vec{\kappa},-\vec{\kappa} ;\mathcal{A}_{\vec{\lambda};R}}^{(2)}$ is cured at any order in $u$ from $\varepsilon$-poles, i.e., such that
\begin{align}
\Gamma_{\vec{\kappa},-\vec{\kappa};\mathcal{A}_{\vec{\lambda};R}}^{(2)}  &  =Z^{\prime}%
\Gamma_{\vec{\kappa},-\vec{\kappa};\mathcal{A}_{\vec{\lambda}}}^{(2)}+Y\Gamma_{\vec{\kappa}
,-\vec{\kappa};\mathcal{B}}^{(2)}\nonumber\\
&  =\frac{1}{2}\left(  \delta_{\vec{\lambda},\vec{\kappa}}+\delta_{\vec{\lambda},-\vec{\kappa}
}\right)  \left(  Z^{\prime}+Z^{\prime}A(n)\right)  +Y
\nonumber\\
&+Z^{\prime}B(n)+Y\left(
A(n)+nB(n)\right)
\end{align}
is finite. It follows from summation over the $n$ nonzero values of $\vec{\lambda}$ that
\begin{equation}
\label{finite2}
\left(  Z^{\prime}+nY\right)  +\left(  Z^{\prime}+nY\right)  \left(
A(n)+nB(n)\right)  = \mbox{finite} \,.
\end{equation}
Comparison of Eqs.~(\ref{finite1}) and (\ref{finite2}) now reveals that
\begin{equation}
\label{resY}
Y   =\frac{Z_{\tau}-Z^\prime}{n}\,.
\end{equation}
Hence, the remaining task in determining the renormalization of $\mathcal{A}_{\vec{\lambda}}$ is to determine $Z^\prime$. To this end, we substitute the result~(\ref{resY}) into Eq.~(\ref{formRenA}) and we then note that
\begin{align}
\mathcal{A}_{\vec{\lambda};R}-\frac{1}{n}\mathcal{B}_{R}  &  =Z^{\prime}\left(
\mathcal{A}_{\vec{\lambda}}-\frac{1}{n}\mathcal{B}\right)
  =\frac{1}{2}\Phi_{\vec{\lambda},-\vec{\lambda};R}\,,
\end{align}
which reveals that
\begin{align}
Z^{\prime}  =Z_{c}\,.
\end{align}
Because $Z_{c}$, $Z_{\tau}$ and $Y$ are at each order in $u$ polynomials in $n$, it follows that in particular for $n\rightarrow0$ that
\begin{subequations}
\begin{align}
\lim_{n\rightarrow0}Z_{c}  &  =\lim_{n\rightarrow0}Z_{\tau}\,,\\
\lim_{n\rightarrow0}Y  &  =\lim_{n\rightarrow0}\left(  \frac{Z_{\tau}-Z_{c}%
}{n}\right) ,
\end{align}
\end{subequations}
whereby we now have completely determined the renormalization of $\mathcal{A}_{\vec{\lambda}}$ in the percolation limit. Note that these results are valid to arbitrary order in perturbation theory. To double-check our reasoning, we also calculated $Z_c$ and $Y$ explicitly in a two-loop calculation yielding
\begin{subequations}
\begin{align}
Z_{c}  &  =1+\frac{u}{\varepsilon}-\left[  \frac{3n-9}{\varepsilon}%
-\frac{5n-47}{12}\right]  \frac{u^{2}}{4\varepsilon}\,,\\
\label{resY2Loop}
Y  &  =-\frac{u}{\varepsilon}+\left[  \frac{5n-11}{\varepsilon}-\frac
{23n-65}{12}\right]  \frac{u^{2}}{4\varepsilon} \,.
\end{align}
\end{subequations}
In conjunction with Eq.~(\ref{2loopResTau}), the results clearly confirm our above general results.

It is worth pointing out that the equality of temperature renormalization and crossover renormalization in the replica limit is a generic feature of all replicated field theories which therefore have a crossover exponent which is strictly equal to 1. For example, this is also true for the $n \to 0$ limit of the $n$-component $\varphi^4$-theory describing linear polymers. Surprisingly, this important general feature has not been much appreciated in the literature, at least as far as we know.

\section{Short distance expansion}
\label{shortDistanceExpansion}
As discussed in Sec.~\ref{probDistDefAndMF}, the distribution function for the nonlinear resistance can be extracted from the two-point correlation function $G(\mathbf{r},\tau,w\Lambda_r(\vec{\lambda}))$, whose scaling form is well known to be given by Eq.~(\ref{ScaleForm2Point}). For treating the limit $|\brm{r}|/R^{\phi_r/\nu} \ll 1$, however, we need to know how the scaling function $\hat{G}(x,y)$ in Eq.~(\ref{ScaleForm2Point}) behaves for small arguments. Naively, one might attempt a Taylor expansion of $\hat{G}(x,y)$ for small arguments which is erroneous, however, because the scaling function contains infra-red (IR) singularities. To treat these IR singularities properly, one has to resort to a short-distance expansion (SDE). For the $\phi^4$-model, this was done a long time ago by Brezin, Amit, and Zinn-Justin \cite{BAZ74}, and Brezin, De\,Dominicis, and Zinn-Justin \cite{BDZ74}. To follow their route, we will briefly review some essentials of the SDE and its basis, the operator product expansion (OPE)~\cite{WKZ69}. Then we will investigate the short distance behavior of $G(\mathbf{r},\tau,w\Lambda_r(\vec{\lambda}))$ by applying the SDE which will involve the operators $\mathcal{A}_{\vec{\lambda}}$ and $\mathcal{B}$ whose renormalization we discussed in Sec.~\ref{representations}. In the following, we will suppress the subscript $R$ on renormalized operators for notational simplicity.

The OPE is based on the fact that a product of two renormalized operators $\mathcal{O}_{\alpha}(\mathbf{r})$ and $\mathcal{O}_{\beta}(\mathbf{r}^\prime)$ (which may be plain or composite fields) at two different but close points can be expanded as~\cite{amit_zinn-justin}
\begin{equation}
\label{generalExp}
\mathcal{O}_{\alpha}(\mathbf{r})\mathcal{O}_{\beta}(\mathbf{r}^{\prime}%
)=\sum_{\gamma}c_{\alpha\beta,\gamma}(\mathbf{r}-\mathbf{r}^{\prime
})\mathcal{O}_{\gamma}((\mathbf{r}+\mathbf{r}^{\prime})/2)
\end{equation}
where the $c_{\alpha\beta,\gamma}(\mathbf{r})$ are pure functions of $\mathbf{r}$. Of course, any decomposition into symmetry representations has to be done such that both sides of this equation remain consistent. The statement~(\ref{generalExp}) is meaningful only when inserted in correlation or vertex functions and only in the critical domain. The same holds true for many of the equations in the remainder of this section, and we ask the reader to keep this in mind when going through this section. If the $\mathcal{O}_{\alpha}(\mathbf{r})$ are eigenoperators of the RG, i.e., if
\begin{equation}
\mathcal{O}_{\alpha}(\ell\mathbf{r})=\ell^{-x_{\alpha}}\mathcal{O}_{\alpha
}(\mathbf{r})\,,
\end{equation}
where $x_{\alpha}$ is the scaling dimension of $\mathcal{O}_{\alpha}(\mathbf{r})$, then the $c_{\alpha\beta,\gamma}(\mathbf{r})$ are given by
\begin{equation}
c_{\alpha\beta,\gamma}(\mathbf{r})\propto\left\vert \mathbf{r}\right\vert
^{x_{\gamma}-x_{\alpha}-x_{\beta}}.
\end{equation}
The higher the scaling dimension of an operator $\mathcal{O}_{\gamma}$ in the expansion (\ref{generalExp}), the less it contributes if $|\brm{r} - \brm{r}^\prime|$ is small. Thus one naturally obtains an expansion for short distances, or SDE, if one orders the operators in the OPE according to their scaling dimension.

In general, the unit operator will be present on the right hand side of Eq.~(\ref{generalExp}). However, if one considers connected correlation functions, as we do, it does not contribute. The leading contributing operators are then those which have at $d_c$ the naive dimension $4$, viz.\ $\mathcal{A}_{\vec{\lambda}} (\brm{R})$ and $\mathcal{B}(\brm{R})$, and thus the leading contributing terms in the SDE of $\tilde{\varphi}_{\vec{\lambda}} ( \brm{r}) \tilde{\varphi}_{-\vec{\lambda}} (\brm{r}^\prime)$ read
\begin{align}
\label{specificFormSDE}
&\tilde{\varphi}_{\vec{\lambda}}(\mathbf{R}+\mathbf{r}/2)\tilde{\varphi}_{-\vec{\lambda}}(\mathbf{R}%
-\mathbf{r}/2)&
\nonumber\\
&=c_{\mathcal{A}}(\mathbf{r})\mathcal{A}_{\vec{\lambda}}(\mathbf{R})+c_{\mathcal{B}}(\mathbf{r})\mathcal{B}(\mathbf{R})+\cdots\,.
\end{align}
All other composite fields such as  $[\nabla\tilde{\varphi}_{\vec{\lambda}}\nabla\tilde{\varphi}_{-\vec{\lambda}}](\mathbf{R})$, $[\tilde{\varphi}_{\vec{\lambda}}\nabla^{2}
\tilde{\varphi}_{-\vec{\lambda}}](\mathbf{R})$, $[\tilde{\varphi}_{\vec{\lambda}}\tilde{\Psi}
_{-\vec{\lambda}}](\mathbf{R})$, and their sum over $\vec{\lambda}$ do at least have naive dimension $6$ at $d_c$. The same holds true for the operator
\begin{align}
\mathcal{C}(\mathbf{R})& := w\, \left[\sum_{ \vec{\lambda}} \Lambda_r(\vec{\lambda})\tilde{\varphi}_{\vec{\lambda}}\tilde{\varphi}_{-\vec{\lambda}} \right] (\mathbf{R}) =: w \, \mathcal{C}^\prime(\mathbf{R}) \,.
\end{align}
Note, that this composite operator must be renormalized as a whole;  its renormalization does not simply follow from the renormalization of its part $[\tilde{\varphi}_{\vec{\lambda}}\tilde{\varphi}_{-\vec{\lambda}}]\sim \mathcal{A}_{\vec{\lambda}}$ in the RRN-limit. This was overlooked in earlier work \cite{WaYo78} and led the wrong result $\phi_r = 1$ for the crossover exponents. However, the operator $\mathcal{C}^{\prime}$ plays a special role due to the double limit $D \to 0$ and $M\to \infty$. When the other dimension $6$ operators such as $\left. \sum\right.  _{ \vec{\lambda}}^{\prime} [\tilde{\varphi}_{\vec{\lambda}}\nabla^{2} \tilde{\varphi}_{-\vec{\lambda}}](\mathbf{R})$ are inserted into Feynman diagrams, they leave the superficial degree of divergence unchanged. When $\mathcal{C}(\mathbf{R})$ is inserted into Feynman diagrams, it reduces the superficial degree of divergence of any of these diagrams by $2$ just as $\mathcal{B}(\brm{R})$ does. Thus, $\mathcal{C}(\mathbf{R})$ should be included on the same footing as $\mathcal{B}(\brm{R})$ as far as making insertions to avoid extra UV singularities if the limit $|\brm{r} - \brm{r}^\prime| \to 0$ is concerned. All other operators whose naive dimension is $6$ or higher, can be neglected in the following.

Having established the specific form of the SDE for the problem at hand, we will now turn to various RGEs that will eventually reveal the short distance behavior of the 2-point correlation function. These RGEs can be set up using standard procedures exploiting the fact that the unrenormalized theory must be independent of the inverse length-scale $\mu$ introduced by renormalization~\cite{amit_zinn-justin}, cf.\ the renormalization scheme~(\ref{reno}). In the RGEs for the current problem we encounter on various occasions the RG differential operator
\begin{align}
D_\mu :=\mu\partial_{\mu}+\beta\partial_{u}+\kappa\tau\partial_{\tau}+\zeta_r w\partial_{w}\,.
\end{align}
which contains the RG-functions
\begin{align}
\label{wilson}
\beta \left( u \right) = \mu \frac{\partial u}{\partial \mu} \bigg|_0 \, ,
\quad
\kappa \left( u \right) = \mu \frac{\partial \ln \tau}{\partial \mu} \bigg|_0 \, ,
\quad
\zeta_r \left( u \right) = \mu \frac{\partial \ln w}{\partial \mu} \bigg|_0 \, ,
\end{align}
and the RG-function
\begin{align}
\label{wilsonGamma}
\gamma \left( u \right) = \mu \frac{\partial }{\partial \mu} \ln Z \bigg|_0 \, ,
\end{align}
where the $|_0$ indicates that bare quantities are kept fixed while taking the derivatives. All these RG-functions are well known and explicit results for them can be straightforwardly reconstructed from the renormalizations and $Z$-factors reviewed in Sec.~\ref{reviewPottsRRN}.

First, let us state the RGEs for the operators $\mathcal{A}_{\vec{\lambda}}$, $\mathcal{B}$ and $\mathcal{C}^\prime$, which we will need as input as we move along:
\begin{subequations}
\label{RGEABC}
\begin{align}
&\left[ D_\mu +\kappa\right]    \,  \mathcal{A}_{\vec{\lambda}}(\mathbf{R})+y\, \mathcal{B}(\mathbf{R})=0\,,
\\
&\left[ D_\mu +\kappa\right]     \, \mathcal{B}(\mathbf{R})=0\,,
\\
&\left[ D_\mu +\zeta_r\right] \,  \mathcal{C}^{\prime}(\mathbf{R})=0\,,
\end{align}
\end{subequations}%
with the RG-function $y(u)$ given by
\begin{align}
y(u) = \mu \frac{\partial }{\partial \mu} Y \bigg|_0 = u-\frac{65}{24}u^{2}+\cdots\, ,
\end{align}
where we used on the right hand side our 2-loop result~(\ref{resY2Loop}) for $Y$. Next, we set up and solve RGEs for the SDE coefficients $c_{\mathcal{A}}$ and $c_{\mathcal{B}}$. From Eq.~(\ref{specificFormSDE}) we learn by using the distributivity of $D_\mu$ and by exploiting the RGEs~(\ref{RGEABC}) that
\begin{align}
0  &  =\left[ D_\mu +\gamma\right]  \left(  \tilde{\varphi}_{\vec{\lambda}}
(\mathbf{R}+\brm{r}/2) \tilde{\varphi}_{-\vec{\lambda}}(\mathbf{R}-\brm{r}/2)\right)
\nonumber\\
&\approx \left[ D_\mu +\gamma\right]  \left(  c_{\mathcal{A}}(\brm{r}%
)\mathcal{A}_{\vec{\lambda}}(\mathbf{R})+c_{\mathcal{B}}(\brm{r})\mathcal{B}%
(\mathbf{R})\right) \nonumber\\
&  =\mathcal{A}_{\vec{\lambda}}(\mathbf{R})\left[ D_\mu +\gamma\right]
c_{\mathcal{A}}(\brm{r})+c_{\mathcal{A}}(\brm{r})D_\mu \mathcal{A}_{\vec{\lambda}}(\mathbf{R})
\nonumber\\
&+\mathcal{B}(\mathbf{R})\left[ D_\mu +\gamma\right]
c_{\mathcal{B}}(\brm{r})+c_{\mathcal{B}}(\brm{r})D_\mu \mathcal{B}%
(\mathbf{R})\nonumber\\
&  =\mathcal{A}_{\vec{\lambda}}(\mathbf{R})\left[ D_\mu +\gamma-\kappa\right]
c_{\mathcal{A}}(\brm{r})
\nonumber\\
&+\mathcal{B}(\mathbf{R})  \left(  \left[
D_\mu +\gamma-\kappa\right]  c_{\mathcal{B}}(\brm{r})-yc_{\mathcal{A}%
}(\brm{r})\right)  \,.
\end{align}
Because the operators $\mathcal{A}_{\vec{\lambda}}$ and $\mathcal{B}$ are linearly independent, this implies that\begin{subequations}
\begin{align}
&  \left[  D_\mu +\gamma-\kappa\right]  c_{\mathcal{A}}(\mathbf{r})=0\,,\\
&  \left[  D_\mu + \gamma-\kappa\right]  c_{\mathcal{B}}(\mathbf{r})-y\, c_{\mathcal{A}}(\mathbf{r})=0\,.
\end{align}
\end{subequations}
At the RG fixed-point $u^\ast$, determined by $\beta (u^\ast) = 0$, these coupled RGEs for $c_{\mathcal{A}}$ and $c_{\mathcal{B}}$ are readily solved with the result
\begin{subequations}
\begin{align}
\label{scaleFormCA}
c_{\mathcal{A}}(\mathbf{r};\mu)  &  =c_{\mathcal{A}}(\mu\mathbf{r})=\left(
\mu\left\vert \mathbf{r}\right\vert \right)  ^{2-\eta-1/\nu}c_{\mathcal{A}}(1)\,,
\\
c_{\mathcal{B}}(\mathbf{r};\mu)  &  =c_{\mathcal{B}}(\mu\mathbf{r})=\left(
\mu\left\vert \mathbf{r}\right\vert \right)  ^{2-\eta-1/\nu}
\nonumber\\
&\times \left[ c_{\mathcal{B}}(1)-y (u^{\ast}) c_{\mathcal{A}}(1)\ln\left(  \mu\left\vert
\mathbf{r}\right\vert \right)  \right]  \,.
\end{align}
\end{subequations}
The additional logarithmic contribution $\ln \left(  \mu\left\vert \mathbf{r}\right\vert \right)$ appears because the operators $\mathcal{A}_{\vec{\lambda}}$ and $\mathcal{B}$ have the same scaling dimension, $d-1/\nu$, in the percolation limit $n\to 0$. $\eta = \gamma^\ast$ and $\nu = 1/(2 - \kappa^\ast)$, where $\gamma^\ast = \gamma (u^\ast)$ and so on, are the usual critical percolation-exponents, cf.~Sec.~\ref{renScalingInt}. For completeness, and because we will use this later on, we also note that the resistance exponent $\phi_r$ is determined by the fixed point values of the RG-functions as $\phi_r = \nu (2-\zeta_r^\ast)$.

Now we are in the position to address the short distance behavior of $G(\mathbf{r},\tau,w\Lambda_r(\vec{\lambda}))$. We must not naively simply plug the expansion~(\ref{specificFormSDE}) into the correlation function~(\ref{defCorr}) because this would generate correlation functions such as $\langle\mathcal{A}_{\vec{\lambda}}(\mathbf{R})\rangle$ which require additional additive renormalizations. This problem can be avoided by taking derivatives with respect to $\tau$ and $w$ which generate insertions of $\mathcal{B}$ and  $\mathcal{C}^\prime$ that reduce the superficial degree of divergence. It turns out that 3-fold differentiation is sufficient to suppress additional $UV$-singularities. In the following, we denote derivatives with respect to $-\tau$ and $-w$ with superscripts $N$ and $M$, e.g.,
\begin{align}
\label{corrWithDer}
&\Lambda_r(\vec{\lambda})^M G^{(N,M)}(\mathbf{r},\tau,w\Lambda_r(\vec{\lambda}))
\nonumber\\
&:=  (-\partial_{\tau})^{N}(-\partial_{w})^{M} G(\mathbf{r},\tau,w\Lambda_r(\vec{\lambda}))
\nonumber \\
&= \left\langle \tilde{\varphi}_{\vec{\lambda}}(\mathbf{r}/2) \tilde{\varphi}_{-\vec{\lambda}
}(-\mathbf{r}/2)\widetilde{\mathcal{B}}(\mathbf{0})^{N}\widetilde{\mathcal{C}%
}^{\prime}(\mathbf{0})^{M}\right\rangle\,,
\end{align}
where
\begin{align}
\widetilde{\mathcal{B}}(\mathbf{q})=\int d^{d}r\,\mathcal{B}(\mathbf{r}) \, \mathrm{e}^{-i\mathbf{q}\cdot\mathbf{r}},
\end{align}
and likewise for $\widetilde{\mathcal{C}}^{\prime}$. Till noted otherwise, we assume that $N+M\geq 3$. Then it is safe to use the SDE~(\ref{specificFormSDE}) in the connected correlation function with insertions~(\ref{corrWithDer}), which leads to
\begin{align}
\label{G2Exp}
&\Lambda_r(\vec{\lambda})^M G^{(N,M)}(\mathbf{r},\tau,w\Lambda_r(\vec{\lambda}))
\nonumber\\
& \approx  c_{\mathcal{A}}(\mathbf{r})\left\langle
\mathcal{A}_{\vec{\lambda}}(\mathbf{0})\widetilde{\mathcal{B}}(\mathbf{0}
)^{N}\widetilde{\mathcal{C}}^{\prime}(\mathbf{0})^{M}\right\rangle
\nonumber\\
&+c_{\mathcal{B}}(\mathbf{r})\left\langle\mathcal{B}(\mathbf{0})\widetilde{\mathcal{B}}%
(\mathbf{0})^{N}\widetilde{\mathcal{C}}^{\prime}(\mathbf{0})^{M}\right\rangle
\,.
\end{align}
This equation makes it transparent that the expansion has the benefit of factorizing the spatial dependence from the dependence on the control parameters $\tau$ and $w$. This feature will be important when it comes to solving the RGE for $G^{(N,M)}$. As pointed out above, this expansion is valid only in the critical domain, i.e., for $\left\vert \tau/\mu^{2}\right\vert ^{\nu}\ll\left\vert \mu\mathbf{r}\right\vert ^{-1}$ and $\left\vert w/\mu^{2}\right\vert ^{\nu/\phi_r}\ll\left\vert \mu\mathbf{r}\right\vert ^{-1}$.

Next, we discuss the 2 contributions to Eq.~(\ref{G2Exp}). The second term on its right hand side vanishes in the limit $n\to 0$. This can be seen as follows. Let us abbreviate $\langle\mathcal{B}(\mathbf{0})\widetilde{\mathcal{B}}(\mathbf{0})^{N}\widetilde{\mathcal{C}}^{\prime}
(\mathbf{0})^{M}\rangle=\sum_{\vec{\lambda}}^\prime\langle
\mathcal{A}_{\vec{\lambda}}(\mathbf{0})\widetilde{\mathcal{B}}(\mathbf{0}
)^{N}\widetilde{\mathcal{C}}^{\prime}(\mathbf{0})^{M}\rangle := \sum_{\vec{\lambda}}^\prime f(\Lambda_r(\vec{\lambda}))$, where we have reinstated the prime on the sums over replicated currents to emphasize that $\vec{\lambda} = \vec{0}$ is excluded from these sums, c.f.\ the comment below Eq.~(\ref{defLambdaAlpha}). Switching to an unrestricted sum, we can rewrite $\sum_{\vec{\lambda}}^\prime f(\Lambda_r(\vec{\lambda}))$ as $\sum_{\vec{\lambda}} f(\Lambda_r(\vec{\lambda})) - f(0)$. By expressing $f(\Lambda_r(\vec{\lambda}))$ in terms of its Laplace transform, switching from discrete to continuous replica currents and then performing the integration over $\vec{\lambda}$ with help of the saddle-point method, we find that $\sum_{\vec{\lambda}} f(\Lambda_r(\vec{\lambda})) = f(0)$ in the limit $D\to 0$. Thus, the second term on the right hand side of Eq.~(\ref{G2Exp}) vanishes in this limit. For discussing the first term on the right hand side of Eq.~(\ref{G2Exp}), we denote
\begin{align}
F_{\mathcal{A}}^{(N,M)}(\tau,w\Lambda_r(\vec{\lambda}),\mu) := \Lambda_r(\vec{\lambda})^{-M} \left\langle
\mathcal{A}_{\vec{\lambda}}(\mathbf{0})\widetilde{\mathcal{B}}(\mathbf{0}
)^{N}\widetilde{\mathcal{C}}^{\prime}(\mathbf{0})^{M}\right\rangle.
\end{align}
An RGE for this connected correlation function is readily derived:
\begin{equation}
\left[  D_\mu +(1+N)\kappa+M\zeta_r\right]  F_{\mathcal{A}}^{(N,M)}(\tau,w\Lambda_r(\vec{\lambda}),\mu)=0\,.
\end{equation}
At the fixed point, it is straightforward to solve this RGE. The result is
\begin{align}
\label{solRGEFA}
F_{\mathcal{A}}^{(N,M)}(\tau,w\Lambda_r(\vec{\lambda}),\mu)&=\ell^{(1+N)\kappa^{\ast}
+M\zeta_r^{\ast}}
\nonumber \\
&\times F_{\mathcal{A}}^{(N,M)}(\ell^{\kappa^{\ast}}\tau
,\ell^{\zeta_r^{\ast}}w\Lambda_r(\vec{\lambda}),\ell\mu).
\end{align}
This solution needs to be augmented with a dimensional analysis,
\begin{align}
\label{dimAnaRGEFA}
F_{\mathcal{A}}^{(N,M)}(\tau,w\Lambda_r(\vec{\lambda}),\mu)&
=\mu^{d-2(N+1)-2M}
\nonumber\\
&\times F_{\mathcal{A}}^{(N,M)}(\tau/\mu^{2},w\Lambda_r(\vec{\lambda}))/\mu^{2},1),
\end{align}
to account for the non-vanishing naive dimensions of its various ingredients. By combining Eqs.~(\ref{solRGEFA}) and (\ref{dimAnaRGEFA}) and by exploiting freedom of choice regarding the flow parameter $\ell$ by setting $\ell = (\tau/\mu^2)^\nu$, we obtain the scaling form
\begin{align}
\label{scaleFormFA}
F_{\mathcal{A}}^{(N,M)}(\tau,w\Lambda_r(\vec{\lambda}),\mu)&=\mu^{d-2(N+1)-2M}
\nonumber\\
&\times \left( \tau/\mu^{2}\right) ^{d\nu-(N+1)-M\phi_r}
\nonumber\\
&\times \hat{F}_{\mathcal{A}}^{(N,M)}\left( \frac{w\Lambda_r(\vec{\lambda})/\mu^{2}}{(\tau/\mu
^{2})^{\phi_r}} \right),
\end{align}
where $\hat{F}_{\mathcal{A}}^{(N,M)}(z)$ is a scaling function. For $z$ around $0$, this scaling function is certainly analytic because the model is for $\tau \neq 0$ away from its critical point, and hence there can be no IR-singularities. The situation is more subtle for $\tau \to 0$ because then the question arises whether a nonzero $w$ makes the system noncritical or not. The most economical way to address this question is to calculate $\hat{F}_{\mathcal{A}}^{(0,0)}(z)$ in dimensional regularization to lowest order beyond mean-field theory. Diagrammatically, this calculation involves the attaching of external legs (propagators) to the usual 1-loop self-energy diagrams followed by an integration over the external momentum (which amounts essentially to a 2-loop calculation). Note that it is sufficient to work for $N=M=0$ in dimensional regularization, at least formally, because this regularization formally sets the otherwise required additive renormalizations equal to zero. We perform this calculation for the physically relevant limits of $r$ discussed above. For $r>0$, we find that $w\neq 0$ is sufficient to make the system noncritical for $\tau =0$. Thus, we expect that
\begin{subequations}
\label{assFormsFNM}
\begin{equation}
\hat{F}_{\mathcal{A}}^{(N,M)}(z)=  K_r \, z^{(d\nu-N-1)/\phi_r-M}
\end{equation}
for $z\rightarrow\infty$, where $K_r$ is some constant. For $r<0$ it appears as if $w\neq 0$ is not sufficient to make the system noncritical for $\tau =0$. However, we encounter the problem that the calculation produces a contribution proportional to $\tau^{-1+\varepsilon}$ that has no counterpart in mean-field theory, i.e., $\hat{F}_{\mathcal{A}}^{(0,0)}(z)$ is not renormalizable with the usual renormalization factors for $r<0$. Currently, we do not know how to overcome this problem or what its physical significance might be. Nevertheless, we expect that 
\begin{equation}
\hat{F}_{\mathcal{A}}^{(N,M)}(z)\propto 1  + K_r \, z^{(d\nu-N-1)/\phi_r-M}
\end{equation}
\end{subequations}
for $z\rightarrow\infty$ if $r<0$. Until the problem with the renormalizability is resolved, this formula should be understood as a proposal, as do our final results for the distribution functions pertaining to $r<0$. We plan to return to the unresolved difficulties for $r<0$ in a future publication.

Now, we move on to the RGE for $G^{(N,M)}$. In its generic form, it reads
\begin{align}
\left[  D_\mu +N\kappa+M\zeta_r \right]  G^{(N,M)}(\brm{r},\tau,w\Lambda_r(\vec{\lambda}),\mu)=0\,.
\end{align}
For solving this equation, it is useful to split it up into 2 parts such that one of the parts collects the derivatives with respect to $\tau$ and $w$ contained in $D_\mu$. At the fixed point $u^\ast$, this procedure leads to
\begin{align}
\label{RGEG2sep}
&\lbrack\mu\partial_{\mu}+\eta+N\kappa^{\ast}+M\zeta_r^{\ast}]G^{(N,M)}
\nonumber\\
&=-[\kappa^{\ast}\tau\partial_{\tau}+\zeta_r^{\ast}w\partial_{w}]G^{(N,M)}%
\nonumber\\
&  =\kappa^{\ast}\tau G^{(N+1,M)}+\zeta_r^{\ast}w\Lambda_r(\vec{\lambda})G^{(N,M+1)}\nonumber\\
& \approx c_{\mathcal{A}}\Big[\kappa^{\ast}\tau F_{\mathcal{A}}^{(N+1,M)}+\zeta_r^{\ast
}w\Lambda_r(\vec{\lambda}) F_{\mathcal{A}}^{(N,M+1)}\Big]\,,
\end{align}
where we have dropped the arguments of $G^{(N,M)}$ and so on for notational simplicity. Note that the part of the equation that has been moved over to the right hand side constitutes the leading terms in an expansion of $G^{(N,M)}$ in powers of $\tau$ and $w\Lambda_r(\vec{\lambda})$. Thus, after the RGE is written in the form of Eq.~(\ref{RGEG2sep}), the SDE is safe for $N+M\geq 2$. For the right hand of the RGE~(\ref{RGEG2sep}), we can draw on the results we established above for $c_{\mathcal{A}}$ and $F_{\mathcal{A}}^{(N,M)}$. We obtain
\begin{align}
&  [\mu\partial_{\mu}+\eta+N\kappa^{\ast}+M\zeta_r^{\ast}]G^{(N,M)}
(\mathbf{r},\tau,w\Lambda_r(\vec{\lambda}),\mu)\nonumber\\
&  \approx  \mu^{d-2(N+M+1)}\left(  \mu\left\vert \mathbf{r}\right\vert \right)
^{2-\eta-1/\nu}\left(  \tau/\mu^{2}\right)  ^{(d\nu-1)-(N+M\phi_r)}
\nonumber\\
&\times \Phi^{(N,M)}\left(\frac{w\Lambda_r(\vec{\lambda})/\mu^{2}}{(\tau/\mu^{2})^{\phi_r }} \right) ,
\end{align}
where $\Phi^{(N,M)} (z)$ is a scaling function which sums up the contributions of the different $\hat{F}_{\mathcal{A}}(z)$, and whose details are unimportant for our main quest. Any solution of the inhomogenous RGE consists, of course, of the general solution of the underlying homogenous equation and a particular solution of the inhomogenous equation. The homogenous equation is independent of $\tau$ and $w$, and its solution is found to be
\begin{align}
\label{solHomRGE}
&  G^{(N,M)}(\mathbf{r},\mu)_{\mathrm{hom}}=\ell^{\eta+N\kappa^{\ast}%
+M\zeta^{\ast}}G^{(N,M)}(\mathbf{r},\ell\mu)_{\mathrm{hom}}\nonumber\\
&  =\mu^{d-2(N+M+1)}\ell^{(d-2+\eta)-(N+M\phi_r)/\nu}G^{(N,M)}(\ell\mu
\mathbf{r},1)_{\mathrm{hom}}
\nonumber\\
&\sim\mu^{d-2(N+1)-2M}\left(  \mu\left\vert
\mathbf{r}\right\vert \right)  ^{-(d-2+\eta)+(N+M\phi_r)/\nu} .
\end{align}
As a particular solution of the inhomogenous equation, we find
\begin{align}
\label{solInhomRGE}
&G^{(N,M)}(\mathbf{r},\tau,w\Lambda_r(\vec{\lambda}),\mu)_{\mathrm{part}} \approx \mu
^{d-2(N+M+1)}\left(  \mu\left\vert \mathbf{r}\right\vert \right)
^{2-\eta-1/\nu}\nonumber\\
&  \times\left( \tau/\mu^2\right)  ^{(d\nu-1)-(N+M\phi_r)}S^{(N,M)}\left(\frac{w\Lambda_r(\vec{\lambda})/\mu^{2}}{(\tau/\mu^{2})^{\phi_r
}} \right),
\end{align}
where $S^{(N,M)}(z)$ is some scaling function determined by a linear first order differential equation containing $\Phi^{(N,M)}(z)$. Now, we are finally in the position to write down the short distance scaling form of the 2-point correlation function via adding Eqs.~(\ref{solHomRGE}) and (\ref{solInhomRGE}) and subsequent integration with respect to $\tau$ and $w$. Introducing the scaling variables
\begin{subequations}
\begin{align}
\label{defScaleVarX}
x&=(\tau/\mu^{2})\left(  \mu\left\vert \mathbf{r}\right\vert \right) ^{1/\nu} ,
\\
\label{defScaleVarY}
y&=(w\Lambda_r(\vec{\lambda})/\mu^{2})\left(  \mu\left\vert \mathbf{r}\right\vert \right)^{\phi_r/\nu} ,
\end{align}
\end{subequations}
we arrive at
\begin{align}
\label{finResG2}
G(\mathbf{r},\tau,w\Lambda_r(\vec{\lambda}),\mu)  &\approx \mu^{d-2}\left(  \mu\left\vert
\mathbf{r}\right\vert \right)  ^{-d+2-\eta}\Big\{A_{0}+A_{1,0}x
\nonumber\\
& +A_{0,1} y +A_{2,0}x^{2}+A_{1,1}xy+A_{0,2}y^{2}
\nonumber\\
& +x^{d\nu-1}S_r(y/x^{\phi_r})\Big\},
\end{align}
where $A_0$ and so on are constants, and where $S_r(z)$ is yet another scaling function that, in principle, can be calculated from $S^{(N,M)}(z)$. Based on Eq.~(\ref{assFormsFNM}), we expect that $S_r(z)$ is analytic near $z=0$, and for $z\gg 1$
\begin{equation}
S_r(z)\propto \theta (-r) + K_r z^{(d\nu-1)/\phi_r}\,.
\end{equation}

We conclude this section by asserting what additional contributions we get if we proceed to next order in the SDE \cite{BDZ74}, i.e., if we include operators that have at $d_c$ a naive dimension of $6$. These operators are $\tau\mathcal{A}_{\vec{\lambda}}$, $w\Lambda_r(\vec{\lambda})\mathcal{A}_{\vec{\lambda}}$, $[\tilde{\varphi}_{\vec{\lambda}}\nabla^{2}\tilde{\varphi}_{-\vec{\lambda}}]$, $[\tilde{\varphi}_{\vec{\lambda}}\tilde{\psi}_{-\vec{\lambda}}]$ and $\nabla^{2}\mathcal{A}_{\vec{\lambda}}$, and their respective scaling exponents are $d$, $d+(\phi_r-1)/\nu$, $d$, $d+\bar{\omega}$ and $d+2-1/\nu$, where $\bar{\omega}= \beta^\prime (u^\ast) = \varepsilon + \cdots$ is the so-called Wegner exponent. The operator $\nabla^{2}\mathcal{A}_{\vec{\lambda}}$ does not contribute to the SDE, because when inserted into correlation functions, it produces zero due to translational invariance: $\langle\nabla^{2}\mathcal{A}_{\vec{\lambda}}(\mathbf{R})\widetilde{\mathcal{B}}(\mathbf{0})^{N}\widetilde{\mathcal{C}}^{\prime}(\mathbf{0})^{M}\rangle=\nabla^{2}\langle\mathcal{A}_{\vec{\lambda}}(\mathbf{R})\widetilde{\mathcal{B}}(\mathbf{0})^{N}\widetilde{\mathcal{C}}^{\prime}(\mathbf{0})^{M}\rangle=0$. Therefore, the upshot here is that the dimension $6$ operators lead to additional analytic contributions in the brackets in Eq.~(\ref{finResG2}) which are of third order in $x$ and $y$, such as $x^3$, $x^2 y$ and so on, as well as additional non-analytic contributions of the form $x^{d\nu}S_{1}(y/x^{\phi_r})$, $x^{d\nu
+\phi_r-1}S_{2}(y/x^{\phi_r})$ and $x^{(d+\bar{\omega})\nu}S_{3}(y/x^{\phi_r})$.

\section{Shape of distribution functions}
\label{shapeDistFuncts}

In this section, we exploit our renormalization group results for 2-point correlation function derived in the previous section to determine the shape of connection probability and the probability distribution for the nonlinear resistance. In the remainder, we will suppress the inverse length scale $\mu$ for notational simplicity.

\subsection{Connection probability}
As reviewed in Sec.~\ref{sec:ConProbAndAvNonlRes}, the connection probability, i.e., the probability for any two sites a distance $|\brm{r}|$ apart being on the same cluster, is proportional to the 2-point correlation function of the RRN evaluated at $w=0$. Hence, the scaling form~(\ref{ScaleForm2Point}) of the 2-point correlation function implies that
\begin{equation}
\chi(\mathbf{r},\tau)=  |\mathbf{r}|^{2-d-\eta}\,  \hat{\chi}\big(\tau |\brm{r}|^{1/\nu}\big)\,,
\end{equation}
with $\hat{\chi}(x)$ being a scaling function that depends on the scaling variable $x$  defined in Eq.~(\ref{defScaleVarX}). In the following we will discuss the asymptotic forms of this scaling function for small and large $x$.

For small $x$, our SDE result~(\ref{finResG2}) implies that
\begin{equation}
\hat{\chi}(x) = A_0 +A_{1}x+A_{2}x^{2}+A_{3}x^{d\nu-1}+\cdots\,,
\end{equation}
where $A_0$ and so on are expansion coefficients. When we go to next order in the SDE, we get additional contributions proportional to $x^{3}$, $x^{d\nu}$ und $x^{(d+\bar{\omega})\nu}$.

Now we turn to large x, i.e., to distances $ |\mathbf{r}|$ that are large compared to the correlation length $\xi\propto |\tau|^{-\nu}$. These distances correspond to small wave vectors $\brm{q}$, for which we can expand the 2-point vertex function $\Gamma_{2}(\mathbf{q})$, i.e., the inverse of the 2-point correlation function, as
\begin{equation}
\Gamma_{2}(\mathbf{q},\tau)=\Gamma_{2}(\mathbf{0,}\tau)\left\{  1+\xi^{2} q^{2}+O(q^{4})\right\} .
\end{equation}
Thus, we obtain by applying the inverse Fourier transformation
\begin{align}
G(\mathbf{r},\tau)&=\frac{1}{\Gamma_{2}(\mathbf{0},\tau)}\int_{\mathbf{q}}%
\frac{\mathrm{e}^{i\mathbf{q}\cdot\mathbf{r}}}{1+\xi^{2}q^{2}}
\nonumber\\
&\propto \frac {1}{\xi^{d}\Gamma_{2}(\mathbf{0,}\tau)}\left(  \frac{\xi}{\left\vert
\mathbf{r}\right\vert }\right)  ^{(d-1)/2}\mathrm{e}^{-\left\vert
\mathbf{r}\right\vert /\xi}\,,
\end{align}
where we have omitted higher order terms. Using $\Gamma_{2}(\mathbf{0,}\tau)\propto\tau^{\gamma}$, where $\gamma=(2-\eta)\nu$, we get
\begin{align}
G(\mathbf{r},\tau)&\propto\frac{\tau^{(d-3+2\eta)\nu/2}}{\left\vert
\mathbf{r}\right\vert ^{(d-1)/2}}\mathrm{e}^{-\left\vert \mathbf{r}\right\vert
/\xi}=\frac{x^{(d-3+2\eta)\nu/2}}{\left\vert \mathbf{r}\right\vert ^{d-2+\eta
}}\mathrm{\exp}\left(  -ax^{\nu}\right),
\end{align}
where $a$ is some constant. This reveals that the large-$x$ behavior of the scaling function of the connection probability is given by
\begin{equation}
\hat{\chi}(x)  = A\, x^{(d-3+2\eta)\nu/2}\, \mathrm{\exp}\left(  -ax^{\nu}\right) + \cdots \, ,
\end{equation}
with constants $A$ and $a$.

\subsection{Resistance and fractal masses}
To establish the general scaling form of the resistance distribution, we recall from Sec.~\ref{probDistDefAndMF} that this distribution can be extracted from the 2-point correlation function via inverse Laplace transformation. Inserting Eq.~(\ref{ScaleForm2Point}) into Eq.~(\ref{inverseTrafoYieldingP}), we readily obtain
\begin{align}
\label{genFormPRes}
P_r(R,\mathbf{r}) &  = \frac{\left\vert \mathbf{r}\right\vert ^{-\phi_r/\nu}}{2\pi i}\int
_{\sigma-i\infty}^{\sigma+i\infty}dy\,\frac{\hat{G}(x,y)}{\hat{G}(x,0)}%
\exp\left( yR/\left\vert \mathbf{r}\right\vert ^{\phi_r/\nu}\right)
\nonumber\\
&  =R^{-1}\, \Xi_r \big(R\left\vert \mathbf{r}\right\vert ^{-\phi_r/\nu},\tau |\mathbf{r}|^{1/\nu}\big)\, ,
\end{align}
with a scaling function $\Xi_r$. In the remainder, we focus on criticality, $\tau =0$, where
\begin{align}
\label{tauZeroForm}
P_r(R,\mathbf{r})   =R^{-1}\, \Omega_r \big(R\left\vert \mathbf{r}\right\vert ^{-\phi_r/\nu}\big) ,
\end{align}
with $\Omega_r (s) =  \Xi_r (s,0)$, and we discuss the asymptotic forms of the distribution when the scaling variable
\begin{align}
s := R|\brm{r}|^{-\phi_r/\nu}
\end{align}
is either small or large.

For large $s$, we draw on our SDE results yielding
\begin{align}
\frac{\hat{G}(x,y)}{\hat{G}(x,0)}&=1+B_{1}y+B_{2}xy+B_{3}y^{2}
\nonumber\\
&+ x^{d\nu-1}\left(S_r(y/x^{\phi_r})-S_r(0)\right)  +\cdots\, .
\end{align}
 This expansion is now inserted into Eq.~(\ref{genFormPRes}). In performing the inverse Laplace transformation, the analytic contributions in $y$ merely produce shorts, $R=0$, in the form of delta functions $\delta(R)$ and their derivatives. What remains reads
\begin{align}
P_r(R,\mathbf{r})  &
=\frac{\left\vert \mathbf{r}\right\vert ^{-\phi_r/\nu}}{2\pi i}\int
_{\sigma-i\infty}^{\sigma+i\infty}dy\,x^{d\nu-1}
\nonumber\\
&\times S_r\left(y/x^{\phi_r}\right)\exp\left(
yR/\left\vert \mathbf{r}\right\vert ^{\phi_r/\nu}\right)
\nonumber\\
&  =\frac{1}{R} \left(  \frac{\left\vert \mathbf{r}\right\vert }{R^{\nu/\phi_r}}\right)
^{d-1/\nu}\frac{1}{2\pi i}\int_{\sigma-i\infty}^{\sigma+i\infty}
dz\,z^{(d\nu-1)/\phi_r}\nonumber\\
&\times \hat{S}_r(\tau R^{1/\phi_r}/z^{1/\phi_r})\,\exp(z)\, ,
\end{align}
where $\hat{S}_r(z)=z^{d\nu-1}S_r(1/z^{\phi_r})$. At the critical point, we obtain in particular
\begin{align}
\label{largeSform}
\Omega_r (s) &= C_{r,1}\, s^{-(d\nu-1)/\phi_r}+C_{r,2}\, s^{-d\nu/\phi_r}
\nonumber\\
&+C_{r,3}\, s^{-(d\nu+\phi_r-1)/\phi_r}+C_{r,4}\, s^{-(d+\bar{\omega})\nu/\phi_r}
\nonumber\\
&+ \cdots \, ,
\end{align}
where $C_{r,1}$ and so on are constants. Here, we included the next higher order terms in the SDE as discussed above.

Finally, we turn to small $s$.  To this end, we expand the 2-point vertex function for small wave vectors,
\begin{align}
&\Gamma_{2}(\mathbf{q};\tau,z)
\nonumber\\
&=\Gamma_{2}(\mathbf{0};
\tau,z)+q^{2}  \partial_{q^{2}}\Gamma_{2}(\mathbf{q};\tau,z)\big\vert _{q^{2}=0}
+O(q^{4})
\nonumber\\
& =z^{\gamma/\phi_r}f_{0}\big(\tau/
z ^{1/\phi_r} \big)\Big\{  1
+q^{2}/z^{2\nu/\phi_r}f_{1}\big(\tau/z^{1/\phi_r} \big)+O(q^{4})\Big\} ,
\end{align}
where we denote $z=w\Lambda_{r}(\vec{\lambda})$ in conformity with our conventions of Sec.~\ref{probDistDefAndMF} and where $f_0$ and $f_1$ are scaling functions. At the critical point, we can replace these functions by constants. By dropping terms that are subdominant for $\left\vert \mathbf{r}\right\vert \gg z^{-\nu/\phi_r}$, we obtain
 \begin{align}
G(\mathbf{r},\tau & =0,z)=\frac{1}{f_{0}z^{\gamma/\phi_r}}%
\int_{\mathbf{q}}\frac{\mathrm{e}^{i\mathbf{q}\cdot\mathbf{r}}}{1+f_{1}%
z^{-2\nu/\phi_r}q^{2}}
\nonumber\\
& \propto\frac{y^{(d-3+2\eta)\nu/2\phi_r}}{\left\vert \mathbf{r}\right\vert
^{d-2+\eta}}\, \mathrm{\exp}\big(  - f_1 y^{\nu/\phi_r}\big) ,
\end{align}
where  $y= |\brm{r}|/z^{\nu/\phi_r}$ in accordance with Eq.~(\ref{defScaleVarY}). Feeding this intermediate result into Eq.~(\ref{genFormPRes}), we obtain
\begin{align}
P_r(R, \mathbf{r})&=\frac{\left\vert \mathbf{r}\right\vert ^{-\phi_r/\nu}}{2\pi
i}\int_{\sigma-i\infty}^{\sigma+i\infty}dy\,y^{(d-3+2\eta)\nu/2\phi_r}
\nonumber\\
&\times \exp\big(  yR/\left\vert \mathbf{r}\right\vert ^{\phi_r/\nu}-f_1 y^{\nu/\phi_r
}\big)  .
\end{align}
The remaining integration can be simplified by substituting
\begin{equation}
y=X^{\phi_r/\nu}v\,,\quad \mbox{with} \quad X\propto \big(|\mathbf{r}|/R^{\nu/\phi_r}\big)^{1/(1-\nu/\phi_r)}\, ,
\end{equation}
which yields
\begin{align}
P_r(R,\mathbf{r})&\propto\frac{X^{(d-1+2\eta)/2}}{R}\frac{1}{2\pi i}\int
_{\sigma-i\infty}^{\sigma+i\infty}dv\,v^{(d-3+2\eta)\nu/2\phi_r}
\nonumber\\
&\times \exp\big( X(v-k v^{\nu/\phi_r})\big) ,
\end{align}
where $k$ is some constant. For large $X$, we may use the saddle-point approximation provided that $\nu/\phi_r<1$. However, $\nu/\phi_r>1$ for $d=2$ and $r=1$ as well as $r=\infty$ since $\nu=4/3$! We finally arrive at
\begin{align}
\label{smallSform}
\Omega_r (s) &=  C_r \, s^{-(d-2+2\eta)\nu/2(\phi_r-\nu)}\exp\big(  -c_r s^{-\nu/(\phi_r-\nu)}\big)
\nonumber \\
&+ \cdots \, ,
\end{align}
where $C_r$ and $c_r$ are constants. As discussed above, it is straightforward to extract from these general results the distributions of the linear total resistance as well ass the distributions of the fractal masses of the red bonds, the backbone and the shortest and longest SAWs by taking the appropriate limit with respect to $r$.

For $P_{\text{SAW}} (L, \brm{r})$, the analysis just presented in conjunction with the method of Ref.~\cite{janssen_stenull_SAW_2007} leads at criticality to the same form as given in Eqs.~(\ref{tauZeroForm}), (\ref{largeSform}) and (\ref{smallSform}). To transcribe the above results to the average SAW, we just need to replace on the right hand side of these equations $R$ by $L$, $\phi_r$ by $\phi_0 = \nu/\nu_{\text{SAW}}$ and the constants $C_{r,1}$ and so on by other constants, say $C_{0,1}$ and so on.

\section{Concluding remarks}
\label{concludingRemarks}
In summary, we calculated the asymptotic forms of the pair-connection probability, the distributions of the total resistance and fractal masses of the backbone, the red bonds and the shortest, the longest and the average self-avoiding walk between any two points on a cluster. Our analysis drew solely on general, structural features of the underlying diagrammatic perturbation theory, and hence our results for the form of the distributions are valid to arbitrary loop order, although the critical exponents featured in these distributions are known only to finite order, of course. The distributions of the transport quantities such as the total resistance or the backbone mass etc.\ are essentially all of the same form. In this sense, these distributions share a high degree of universality.

As far as future directions are concerned, we think it is worthwhile to study the renormalizability-problem for $r<0$ discussed in Sec.~\ref{shortDistanceExpansion} and its implications in more detail. It would be interesting to have field theoretic results for distribution functions in directed and dynamic percolation. Also, it would be interesting to see experimental or simulation results for distribution functions in any kind of percolation that might be compared to our results presented here or to future results. We hope that our work stimulates further interest in this subject.

\begin{acknowledgments}
  This work was supported in part (O.S.) by the National Science Foundation under
  grant No.\ DMR 0804900.
\end{acknowledgments}


\begin{thebibliography}{99}
\bibitem{footnote:meanFieldDist}
The calculation is simplest, if one carries out the inverse Laplace transformation before the Fourier transformation.
\end{thebibliography}
\end{document}